\documentclass{emulateapj}

\newcommand{\teff}{$T_{\rm eff}$ }
\newcommand{\tsin}{$T_{\rm eff}$}
\newcommand{\tef}{T_\mathrm{eff}}

\slugcomment{Submitted to ApJ}

\shorttitle{Dissecting 18 Sco with high precision spectroscopy}
\shortauthors{Mel\'{e}ndez et al.}

\begin{document}

\title{18 Sco: a solar twin rich in refractory and neutron-capture elements. Implications for chemical tagging. \altaffilmark{*}}

\author{
Jorge Mel\'{e}ndez\altaffilmark{1},
Iv\'an Ram{\'{\i}}rez\altaffilmark{2},
Amanda I. Karakas\altaffilmark{3},
David Yong\altaffilmark{3},
TalaWanda R. Monroe\altaffilmark{1},
Megan Bedell\altaffilmark{4},
Maria Bergemann\altaffilmark{5},
Martin Asplund\altaffilmark{3},
Marcelo Tucci Maia\altaffilmark{1},
Jacob Bean\altaffilmark{4},
Jos\'e-Dias do Nascimento\altaffilmark{6,7},
Michael Bazot\altaffilmark{8},
Alan Alves-Brito\altaffilmark{9},
Fabr{\'{\i}}cio C. Freitas\altaffilmark{1},
Matthieu Castro\altaffilmark{6}
}
\email{jorge.melendez@iag.usp.br}
\altaffiltext{1}{Departamento de Astronomia do IAG/USP, Universidade de S\~ao Paulo, Rua do Mat\~ao 1226, 
05508-900 S\~ao Paulo, SP, Brazil}
\altaffiltext{2}{McDonald Observatory and Department of Astronomy,
University of Texas at Austin, USA}
\altaffiltext{3}{Research School of Astronomy and Astrophysics,
The Australian National University, Cotter Road, Weston, ACT 2611, Australia}
\altaffiltext{4}{Department of Astronomy and Astrophysics, University of Chicago, 5640 S. Ellis Ave., Chicago, IL 60637, USA}
\altaffiltext{5}{Institute of Astronomy, University of Cambridge, Madingley Road, CB3 0HA, Cambridge, UK}
\altaffiltext{6}{Departamento de F{\'{\i}}sica Te\'orica e Experimental, Universidade Federal do Rio Grande do Norte, 59072-970 Natal, RN, Brazil}
\altaffiltext{7}{Harvard-Smithsonian Center for Astrophysics, Cambridge, Massachusetts 02138, USA}
\altaffiltext{8}{Centro de Astrof\'{i}sica da Universidade do Porto, Rua das Estrelas, 4150-762 Porto, Portugal}
\altaffiltext{9}{Instituto de Fisica, Universidade Federal do Rio Grande do Sul, Av. Bento Goncalves 9500, Porto Alegre, RS, Brazil}
\altaffiltext{*}{Based on observations obtained at the
European Southern Observatory (ESO) Very Large Telescope (VLT) at Paranal Observatory and
at the 3.6m telescope at La Silla Observatory, Chile (observing programs 083.D-0871 and 188.C-0265) and 
at the W.M. Keck Observatory, which is operated jointly by Caltech, the University of California and NASA.}


%\altaffiltext{1}{Visiting Astronomer, Magellan Telescopes, Las Campanas Observatory, Chile,
%and W.M. Keck Observatory, Manua Kea, Hawaii}


\begin{abstract}
We study with unprecedented detail the chemical composition and stellar parameters of the 
solar twin 18 Sco in a strictly differential sense relative to the Sun. 
Our study is mainly based on high resolution (R~$\sim$~110 000) high S/N (800-1000) 
VLT UVES spectra, which allow us to achieve a precision of about 0.005 dex in differential abundances.
The effective temperature and surface gravity of 18 Sco are \teff = 5823$\pm$6 K and log $g$ = 4.45$\pm$0.02 dex, 
i.e., 18 Sco is 46$\pm$6~K hotter than the Sun and log~$g$ is 0.01$\pm$0.02~dex higher.
Its metallicity is [Fe/H] = 0.054$\pm$0.005 dex and its microturbulence velocity is +0.02$\pm$0.01 km~s$^{-1}$ 
higher than solar.
Our precise stellar parameters and differential isochrone analysis show that 18 Sco has a mass 
of 1.04$\pm$0.02M$_\odot$ and that it is $\sim$1.6~Gyr younger than the Sun.
We use precise HARPS radial velocities to search for planets, but none were detected.
The chemical abundance pattern of 18 Sco displays a clear trend with condensation temperature,
showing thus higher abundances of refractories in 18 Sco than in the Sun.
Intriguingly, there are enhancements in the neutron-capture elements relative to the Sun.
Despite the small element-to-element abundance differences among nearby n-capture elements 
($\sim$0.02 dex), we successfully reproduce the $r$- process pattern in the solar system.
This is independent evidence for the universality of the $r$-process.
Our results have important implications for chemical tagging in our Galaxy 
and nucleosynthesis in general.
\end{abstract}



\section{Introduction}

Solar twins are stars nearly indistinguishable from the Sun \citep{cay96}. 
The star 18 Sco was first identified as a solar twin by
\cite{por97}. This star has great importance because it is the brightest (V = 5.51, Ram\'{i}rez et al. 2012)  
and closest (13.9 pc) solar twin \citep{por97,sou04,tak07,dat12,dat14,por14}, thus it can be studied 
through a variety of techniques. Moreover, 18 Sco has a declination of $-$8 $^\circ$,
hence being observable from both the Northern and Southern hemispheres.

Besides the many recent high resolution chemical abundance studies 
\citep[e.g.][]{lh05,mel07,nev09,ram09a,tak09,gh10,das12,mon13}, 18 Sco has been observed for 
chromospheric activity \citep[e.g.][]{hal07}, magnetic fields \citep{pet08}, 
debris disks \citep[e.g.][]{tri08}, companions through high resolution imaging \citep[e.g.][]{tan10}, 
granulation \citep{ram09b}, seismology \citep{baz11,baz12} and 
interferometry \citep{baz11,boy12}. In addition, different techniques can be
combined to obtain further insights on the fundamental properties of this
solar twin \citep{li12}.

Most previous abundance studies on 18 Sco 
report a somewhat enhanced (about 10-15\%) iron abundance and a Li content 
about 3-4 times higher than solar, but otherwise about 
solar abundance ratios for other elements \citep[e.g.][]{por97,mel07,tak09}, except 
for some recent high precision studies on 18 Sco \citep[e.g.,][]{mel09,ram09a,mon13}, which show a clear trend 
with condensation temperature.
The situation regarding the heavy elements
is less clear, with \cite{por97} reporting a slight excess in the elements 
heavier than Sr, but the recent study by \cite{mis13} finding about solar ratios 
for the neutron-capture elements in 18 Sco. Instead,
\cite{das12} found a clear enhancement in
Sr, Ba, Nd and Sm, but solar ratios for Y and Ce.
To further complicate the situation, \cite{gh10} found a solar ratio for Nd
(an element that was found enhanced by da Silva et al. 2012), but Eu and Zr
showed the largest enrichment. The worst cases are Zr and Nd, with a spread of 0.15 dex and 0.14 dex,
respectively, and individual values of [Zr/Fe] = -0.05, +0.06, +0.10, and [Nd/Fe] = -0.01, 0.13, 0.00,
according to \cite{mis13}, \cite{das12} and \cite{gh10}, respectively.

In order to better understand the likely chemical differences between 18 Sco and the Sun, 
and to clarify the situation regarding the heavy elements (Z $>$ 30), we perform a 
highly precise abundance analysis of 18 Sco for 38 chemical elements, 
thus being the most complete and precise abundance study to date on the chemical composition of 18 Sco. 
Additionally, our precise stellar parameters will be used in a forthcoming paper to better 
constrain the fundamental properties of 18 Sco in synergy with other techniques such as asteroseismology
(Bazot et al., in preparation). 

Our work is also relevant regarding ``chemical tagging'' \citep{fre02}, 
that aims to reconstruct the build up of our Galaxy by identifying stars with a
common origin. Although the dynamical information about their origin could have been lost, 
the chemical information should be preserved. In this context, the disentangling of the complex
abundance pattern of 18 Sco can help us to assess which elements should be targeted
for chemical tagging.

The analysis is mainly based on UV-optical spectra acquired with the UVES spectrograph 
at the VLT and complemented with optical spectra taken with HIRES at Keck.

\section{Observations}

In order to cover a wide spectral range, we observed 18 Sco and the Sun 
(using solar reflected light from asteroids) in different spectrograph configurations. 
Both 18 Sco and the reference solar spectrum were acquired in the same observing runs
and using identical setups.
We obtained visitor mode observations with the UVES spectrograph at the VLT (August 30, 2009) 
and with the HIRES spectrograph at Keck (June 16, 2005), covering with both datasets 
the UV/optical/near-IR spectrum from 306 to 1020 nm. 

The UVES observations were taken in dichroic mode, obtaining thus simultaneous 
UV (blue arm) + optical (red arm) coverage with the standard settings 346 nm + 580 nm, 
and another set of observations with the standard 346-nm setting plus a non-standard 
setting centered at 830 nm. With this configuration we achieved a high S/N in the UV, 
as the 346nm setting (306-387 nm) was covered in both setups. 
The 580-nm standard setting covered the optical (480 - 682 nm) region and our 
830-nm setting included the red region (642 - 1020 nm). Notice that our non-standard 
setting at 830 nm was chosen to overlap the 580 nm setting in the 
642 - 682 nm interval, so that a higher S/N was achieved around 670 nm in order to 
measure lithium with extremely high precision \citep[e.g.,][]{mon13}. 

The bulk of the analysis is based on the UVES optical spectra obtained in the red arm 
covering the 480 - 1020 nm region. The observations with the red arm were obtained using 
the 0.3 arcsec slit, achieving a high resolving power (R = 110 000) and a very high S/N  
(typical S/N $\sim$ 800 pixel$^{-1}$, and around the Li feature S/N $\sim$ 1000). 
Some elements showing lines in the UV were studied with the UVES blue arm using 
the 346 nm setup (306 - 387 nm) with a slit of 0.6 arcsec, resulting in high resolution 
(R = 65 000) high S/N ($\sim$ 600 at 350 nm) spectra. The asteroid Juno was employed to obtain 
our reference solar spectrum for the UVES observations, and similar S/N were achieved both for 18 Sco and Juno. 

The spectral regions 387-480 nm and 577-585nm are missing in our UVES data, 
thus, for them we employed high resolution (R = 100 000) high S/N ($\sim$ 400) 
spectra obtained with the HIRES spectrograph at Keck, that covers the optical spectra (388 - 800 nm) 
using a mosaic of 3 CCDs. The asteroid Ceres was used to obtain a solar spectrum for our HIRES observations. 
In \cite{mel12} we made a detailed comparison of both UVES and HIRES observations of
18 sco relative to the Sun, and concluded that there is an excellent agreement between both 
datasets, resulting in negligible abundances differences (mean difference of 0.002$\pm$0.001 dex 
and element-to-element scatter $\sigma$ = 0.005 dex), thus we complement our UVES equivalent 
width measurements with HIRES data when needed. 

\begin{figure}
\epsscale{1.1}
\plotone{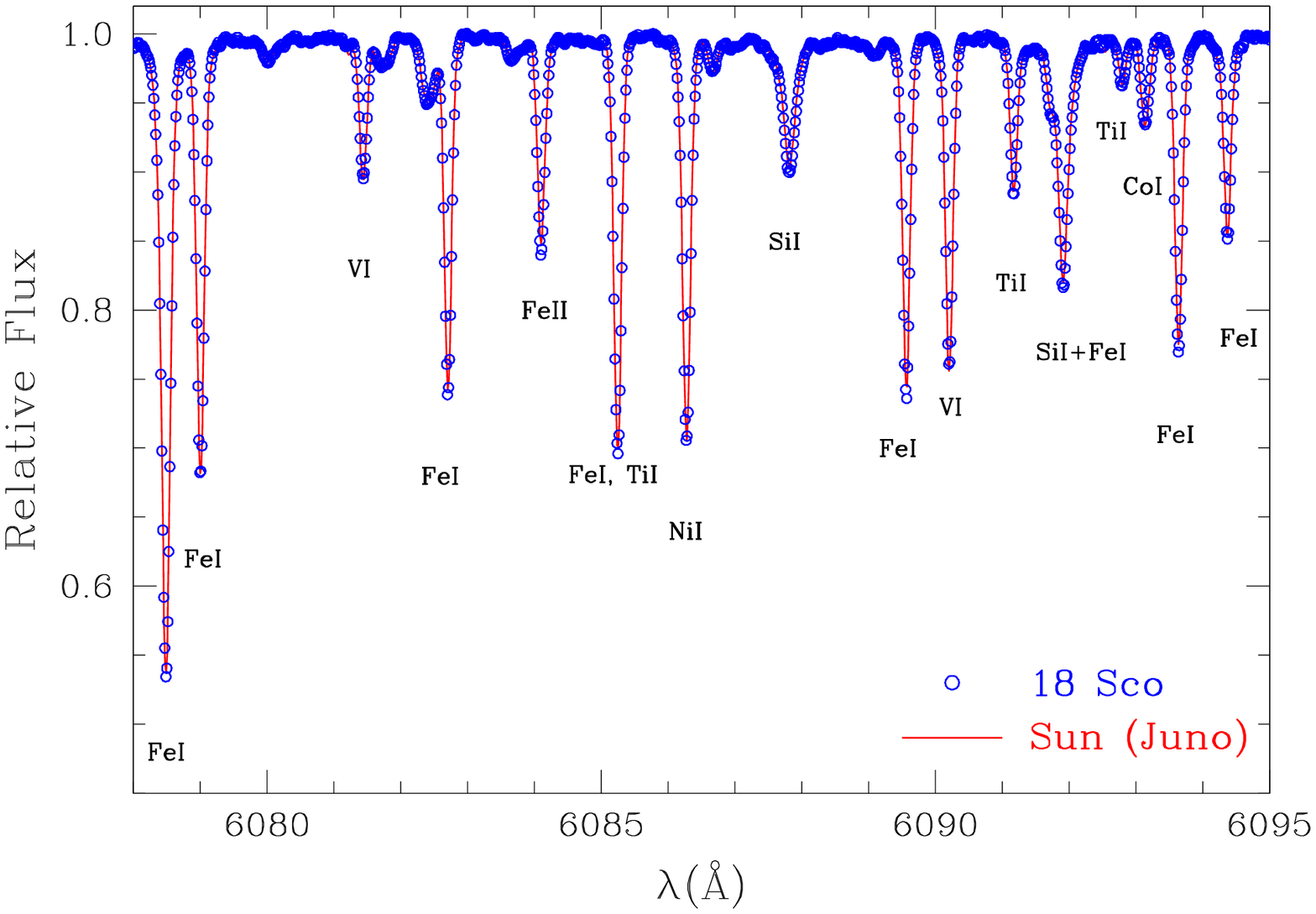}
\caption{UVES spectra of 18 Sco and the Sun in the 6078-6095 \AA\ region. 
Albeit both stars show similar spectra, their different chemical compositions
can be revealed through careful measurements.
\label{compara}}
\end{figure}

\begin{figure}
\epsscale{1.1}
\plotone{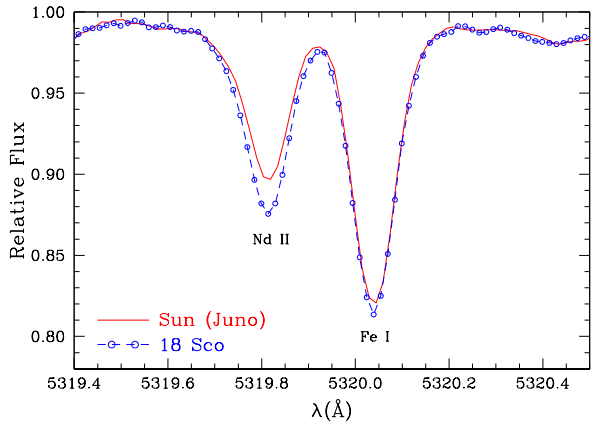}
\caption{UVES spectra of 18 Sco and the Sun around 5320 \AA, 
showing a clear enhancement in the n-capture element Nd in 18 Sco,
relative to the Sun.
\label{comparand}}
\end{figure}

Data reductions of the UVES and HIRES spectra are described in \cite{mon13} and \cite{mel12}, respectively.
A comparison of the reduced UVES spectra of 18 Sco and the Sun is shown around 6085 \AA\ (Fig. \ref{compara}) 
and 5320 \AA\ (Fig. \ref{comparand}).
As can be seen in Fig. \ref{compara}, overall the spectrum of 18 Sco is very similar to the Sun's, 
as expected for a solar twin, yet, when a closer look is taken for the lines of neutron-capture 
elements, a clear enhancement is seen in 18 Sco, as shown for example in Fig. \ref{comparand} for the 5319.8 \AA\ \ion{Nd}{2} line.

\section{Abundance analysis}

The abundance analysis closely follows our recent high precision studies on the solar twins 
HIP 56948 \citep{mel12} and HIP 102152 \citep{mon13}. We measured the 
equivalent widths (EW) automatically with ARES \citep{sou07} for lines with EW larger than 10 m\AA. 
Weaker lines, as well as species with less than 5 lines available, were measured by hand using IRAF. 
In further iterations the outliers resulting from the automatic EW measurements are carefully measured by hand, 
making sure that exactly the same criteria are used in the measurements of 18 Sco and the Sun, 
i.e., for each line we choose the same continuum definition and the same interval was selected 
to fit a gaussian profile. The main difference in relation to \cite{mel12} and \cite{mon13}, is that 
we have significantly expanded our line list to obtain more precise results and also to include many heavy elements.
For example, in \cite{mel12} only 54 iron (42 \ion{Fe}{1}, 12 \ion{Fe}{2}) and 12 chromium (7 \ion{Cr}{1}, 5 \ion{Cr}{2}) lines
were included, while in the present work 98 iron (86 \ion{Fe}{1}, 12 \ion{Fe}{2}) and 21 chromium (14 \ion{Cr}{1}, 7 \ion{Cr}{2}) 
lines are used. In comparison to \cite{mon13}, we have 87 additional lines, many of which
were included to study the neutron-capture elements.

Since our abundances were estimated from EW, we selected mostly clean lines. 
For example, the oxygen abundance was estimated using the 
clean \ion{O}{1} triplet at 777nm rather than the blended forbidden [\ion{O}{1}]
line at 630nm. When necessary we used lines somewhat affected by blending, 
measuring them by using multiple gaussians with the deblend option of the task splot in IRAF.
The list of lines with the differential equivalent width measurements is presented in Table \ref{EW},
except for nitrogen and lithium, that were analysed by spectral synthesis of a NH feature at
340nm and the \ion{Li}{1} feature at 670.7nm, respectively.

We obtain both stellar parameters and elemental abundances through a differential
line-by-line analysis \citep[e.g.,][]{mel12,mon13,ram11,ram14a},
using Kurucz ODFNEW model atmospheres \citep{cas04} and the 2002 version of MOOG \citep{sne73}.
For the Sun we fixed \teff = 5777 K and log $g$ = 4.44 \citep{cox00} and
as initial guess we used a microturbulence velocity of v$_t$ = 0.86 km s$^{-1}$ \citep{mon13}. 
Solar abundances were then computed and the final solar microturbulence was 
found by imposing no trend between the abundances of \ion{Fe}{1} lines and 
reduced equivalent width (EW$_r$ = EW/$\lambda$). We found v$_t^\odot$ = 0.88 km s$^{-1}$
and used this value and the above \teff and log $g$ to compute the reference solar abundances ($A_i^\odot$).

Next, adopting as initial guess for 18 Sco the solar stellar parameters,
we computed abundances for 18 Sco ($A_i^*$), and then
{\it differential} abundances for each line $i$, 

\begin{equation}
\delta A_i = A_i^* - A_i^\odot.
\end{equation}

The effective temperature is found by imposing the differential excitation equilibrium of $\delta A_i$ for \ion{Fe}{1} lines:

\begin{equation}
 d(\delta A_i^{\rm FeI}) / d(\chi_{\rm exc}) = 0 ,
\end{equation}

\noindent while the differential ionization equilibrium of \ion{Fe}{1} and \ion{Fe}{2} lines was used to 
determine the surface gravity:

\begin{equation}
< \delta A_i^{\rm FeII}> - < \delta A_i^{\rm FeI} > \; = \; 0 .
\end{equation}

The microturbulence velocity, v$_t$, was obtained when the differential \ion{Fe}{1} abundances
$\delta A_i^{\rm FeI}$ showed no dependence with the logarithm of the
reduced equivalent width:

\begin{equation}
 d(\delta A_i^{\rm FeI}) / d(\log_{10} (EW_r)) = 0 .
\end{equation}

\begin{figure}
\epsscale{1.1}
\plotone{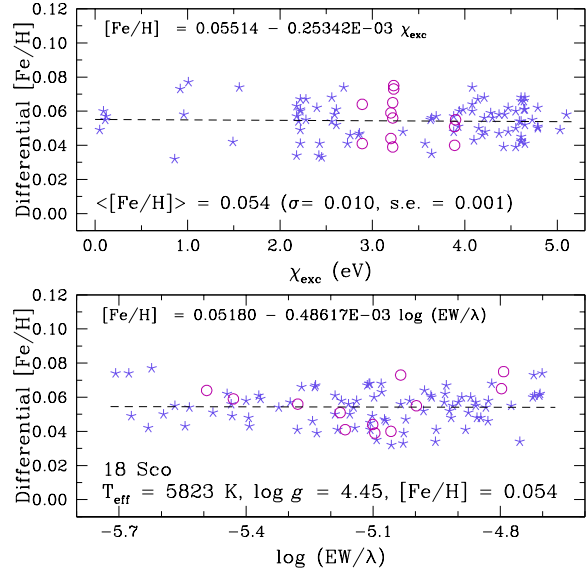}
\caption{Differential iron abundances versus excitation potential (top panel)
and reduced equivalent width (bottom panel) of \ion{Fe}{1} (stars)
and \ion{Fe}{2} (circles) lines. The dashed lines show fits to the \ion{Fe}{1} lines.
The solution is achieved when the slopes are equal or smaller than one third
of the error bar in the slopes, and when the mean abundance of \ion{Fe}{2} agrees with
the mean abundance of \ion{Fe}{1} within one third of the combined error bar.
\label{texc}}
\end{figure}

The spectroscopic solution is reached when the three conditions 
above (eqs. 2-4) are satisfied simultaneously to better than $\sim$1/3 the
error bars in the slopes of eqs. 2 and 4, and better than $\sim$1/3 of the error bar of the
quadratic sum of the error bars of \ion{Fe}{1} and \ion{Fe}{2} lines for eq. 3.\footnote{When it was difficult 
to achieve convergence using the criteria of 1/3 of the error bars, 
we relaxed our criteria to solutions better than 1/2 of the error bars.}
We preferred to adopt these convergence criteria based on the observational error bars rather than using fixed values.
Also, the metallicity obtained from the iron lines must be the same as that of the 
input model atmosphere (within 0.01 dex). 

We emphasise that our iron line list was built to minimize potential 
correlations between the atmospheric parameters, by including lines of
different line strengths at a given excitation potential, and by
having, inasmuch as possible, a homogeneous distribution of lines with
excitation potential. Besides, we keep in our line list only iron lines 
that could be reliably measured at high precision at our spectral resolution.

The differential spectroscopic equilibrium (Fig. \ref{texc})\footnote{Notice that 
the strongest iron lines do not have a significant impact in our final stellar parameters.
If we remove the lines with log (EW/$\lambda$) $> -4.8$, the spectroscopic equilibrium 
would be preserved for \teff and log $g$, but only at the 1-$\sigma$ level for the microturbulence.
The spectroscopic equilibrium would be fully recovered by changing $v_t$ by only $-0.01$ km s$^{-1}$,
without any impact on both \teff and log $g$. The resulting [Fe/H] would be only 0.002 dex higher.
}
results in the following stellar parameters: \teff = 5823$\pm$6 K (46$\pm$6~K hotter than the Sun), 
log $g$ = 4.45$\pm$0.02 dex (+0.01$\pm$0.02~dex above the Sun), 
[Fe/H] = 0.054$\pm$0.005 dex, and $\Delta v_t$=+0.02$\pm$0.01 km~s$^{-1}$ higher than solar.
The above errors include both the measurement uncertainties (from the errors in
the slopes and the errors in the iron abundances), and 
the degeneracies in the stellar parameters, 
by estimating how the error in a given stellar parameter affects the uncertainty
in the others. For example, besides the uncertainty in log $g$
due to the errors in \ion{Fe}{1} and \ion{Fe}{2}, we estimated systematic errors in 
log $g$ due to changes in \ion{Fe}{2} - \ion{Fe}{1} owing to the uncertainties in \tsin, $v_t$ and [Fe/H].

\begin{figure}
\epsscale{1.1}
\plotone{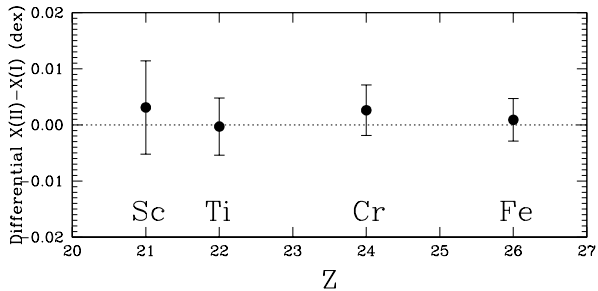}
\caption{Singly-ionized minus neutral differential abundances of Fe, Cr, Ti and Sc. 
The surface gravity found by the ionization equilibrium of iron also
satisfies, within the error bars, the ionization equilibrium of Sc, Ti and Cr.
\label{ion}}
\end{figure}

In \cite{mel12} we found that the small differential NLTE corrections to \ion{Fe}{1} lines 
in the solar twin HIP 56948 do not affect the excitation temperature derived in LTE. Here we 
also computed NLTE corrections for \ion{Fe}{1} as in \cite{ber12}. Again, the differential NLTE corrections
are so small that they do not have any impact on our spectroscopic solution.
Notice that, within the error bars, the differential ionization equilibrium is 
satisfied also by Cr, Ti and Sc, as shown in Fig. \ref{ion}. This good agreement among
different species reinforces our results.

Our stellar parameters are in excellent agreement with those independently determined
by \cite{mon13}, who found a \teff only 1 K hotter, exactly the same log $g$ and $v_t$
and [Fe/H] only 0.001 dex higher,
and by \cite{tak09}, who determined \teff = 5826$\pm$5 K, log $g$ = 4.45$\pm$0.01 dex, and
[Fe/H] = 0.06$\pm$0.01 dex, using high resolution ($R$ = 90 000) high S/N ($\sim$1000 at 600 nm)
HDS/Subaru spectra. Our results are also in firm agreement with stellar
parameters recently determined by \cite{ram14b} using several
high resolution (R = 65000 - 83 000) high S/N (= 400) spectra taken with the MIKE
spectrograph at the Magellan telescope,  \teff = 5816$\pm$4 K, 
log $g$ = 4.45$\pm$0.01 dex, and [Fe/H] = 0.053$\pm$0.003.
Also, there is a good agreement with other results found in the literature, as well as 
an exceptional accord with their weighted mean value, \teff = 5822$\pm$4 K, 
log $g$ = 4.45$\pm$0.01 dex, and [Fe/H] = 0.053$\pm$0.004, as shown in Table \ref{parameters}.

We took hyperfine structure (HFS) into account for 11 elements. 
The calculation is performed including HFS for each individual line and then 
a differential line-by-line analysis is performed. Also, isotopic splitting was
taken into account for the heavier elements.
For V, Mn, Ag, Ba, La, Pr the combined HFS+isotopic splitting is 
a minor differential correction ($\leq$ 0.002 dex), but for Co and Cu the 
differential correction amounts to 0.004 dex, 
for Y the correction is 0.005 dex, and for Yb it is very large at 0.023 dex. The most 
dramatic case is for Eu, for which neglecting the corrections would result 
in an error of 0.155 dex in the differential abundances. 

As shown in \cite{mel12} and \cite{mon13}, differential NLTE effects in 
solar twins relative to the Sun are minor. Here,
we consider differential NLTE corrections for elements showing the largest
differential corrections in our previous works, Mn \citep{ber08} and
Cr \citep{ber10}, but the largest differential correction is only 0.003 dex for Mn.
As mentioned above, differential NLTE effects on Fe \citep{ber12} were also
estimated to check for potential systematics in our differential stellar parameters,
but there is no impact in our solutions.

Our differential abundances (which are based on EW measured by
J. Mel\'endez) are in excellent agreement with those obtained using
an independent set of EW measurements in 18 Sco by Monroe et al. (2013), 
with an average difference of 0.002 dex (this work - Monroe et al.) and 
an element-to-element scatter of only 0.005 dex. Another
independent set of EW measurements obtained by M. Tucci Maia
(that were obtained fully by hand, unlike the measurements done by
J.M and T.M., which used ARES first and then re-measured the outliers by hand), 
results in abundances with a difference from our work of 0.002 dex 
and scatter of only 0.004 dex.
These comparisons, and our previous testing in \cite{mel12}, for
which we obtained an element-to-element scatter of $\sigma$ =0.005 dex,
in the similarity of HIRES and UVES abundances of 18 Sco minus the Sun, suggest that 
careful differential measurements can achieve a precision of about 0.005 dex in differential abundances.

The measurement errors are adopted as the standard error of the differential abundances, 
except for elements with just a single line, in which case we adopted as observational error 
the standard deviation of five differential EW measurements performed with somewhat different criteria.
The typical measurement uncertainties in the differential abundances of the
lighter elements (Z $\leq$ 30) are about 0.004 dex, in good agreement with
the measurement errors discussed above. 
Including the systematic errors
due to uncertainties in the stellar parameters, the total error is about 0.007 dex.
The differential abundances for each element and their errors are given in Table \ref{abun}.

\section{Mass, age, rotation and lithium}

We determine the mass and age of 18 Sco using our precise stellar parameters 
(\teff = 5823$\pm$6 K, log $g$ = 4.45$\pm$0.02 dex, [Fe/H] = 0.054$\pm$0.005 dex) 
and Yonsei-Yale isochrones \citep{yi01,kim02,dem04}. The method, which uses the 
stellar parameters, their error bars, and probability distribution functions,
is described in detail in \cite{mel12} and \cite{cha12}. The method was calibrated 
to reproduce the solar values, as described in \cite{mel12}.

\begin{figure}
\epsscale{1.1}
\plotone{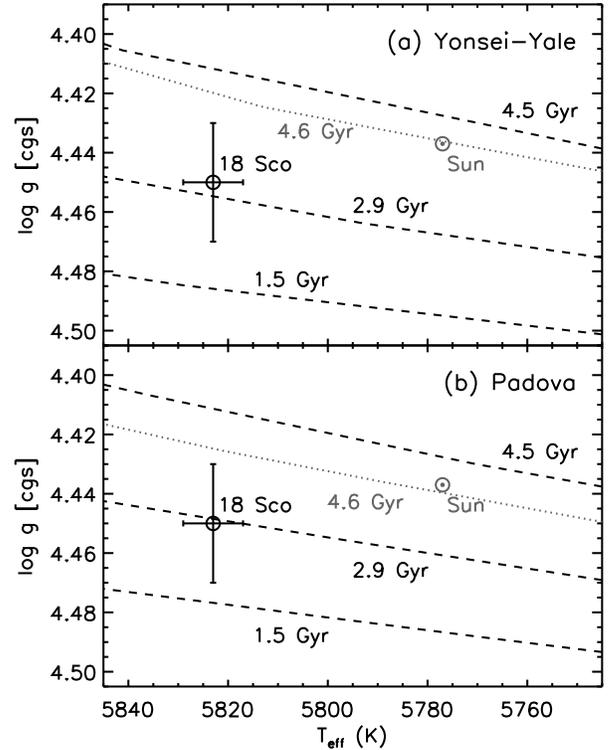}
\caption{Location of the Sun and 18 Sco on the HR diagram. 
(a): Yonsei-Yale isochrones of [Fe/H]=+0.05 (the metallicity of 18 Sco) 
and age=1.5, 2.9, and 4.5 Gyr are shown (dashed lines) along with a solar metallicity 
isochrone of solar age (dotted line). The high precision of our derived stellar parameters 
for 18 Sco allows us to infer a reasonable estimate of its age from the theoretical isochrones, 
even though they are densely packed in this main-sequence region.
(b): as above for Padova isochrones, showing consistent results for
the relative ages between 18 Sco and the Sun.
\label{age}}
\end{figure}

We obtain an age of 2.9$_{-1.0}^{+1.1}$ Gyr for 18 Sco, i.e., 1.6 Gyr younger than the Sun, 
for which we derived an age of 4.5 Gyr using the same set of isochrones \citep{mel12}.
Stellar ages can be well-constrained from isochrones, provided that stellar parameters are known 
with extreme precision, as shown in Fig. \ref{age}a, where we compare our stellar parameters and error bars
to the Yonsei-Yale isochrones. We show in Fig. \ref{age}b that Padova isochrones\footnote{http://stev.oapd.inaf.it/cgi-bin/cmd} \citep{bre12} are compatible
with the relative ages between the Sun and 18 Sco obtained from Yonsei-Yale isochrones.
Our age agrees well with the value found by
\cite{don09}, 2.89$_{-0.81}^{+1.09}$ Gyr, using lithium abundances and stellar parameters.
Within the error bars, our age also agrees with that determined by \cite{li12}, 3.66$_{-0.50}^{+0.44}$ Gyr,
using different constraints (stellar parameters, lithium abundance, rotation and average
large frequency separation). 
Notice, however, that the adopted stellar parameters by
\cite{li12} are not as precise as those reported in this work. We are currently
modeling our HARPS seismic observations of 18 Sco to obtain even better constraints on 
its age.

Following \cite{mel12}, we obtain $v$~sin~$i_{\rm 18 Sco}$/$v$~sin~$i_\odot$ = 1.069$\pm$0.029.
Adopting $v$~sin~$i_\odot$ = 1.90 km s$^{-1}$ for the Sun \citep{bru84,saa97}, 
this implies $v$~sin~$i_{\rm 18 Sco}$ = 2.03$\pm$0.05 km s$^{-1}$. The higher rotation rate in
18 Sco is compatible with its younger age. Fortunately, the rotation period has been determined
for this star, $P$ = 22.7$\pm$0.5 days \citep{pet08}, resulting in a 
rotational age of 3.36$\pm$0.52 Gyr using the rotation-age relationship given in \cite{bar07}.
This value is in good agreement with our derived isochronal age (2.9$_{-1.0}^{+1.1}$ Gyr).

The mass derived here is 1.04$\pm$0.02 M$_\odot$, which agrees well with the
detailed study of \cite{li12}, who reported 1.045 M$_\odot$
%1.045$\pm$0.035$\pm$0.005 M$_\odot$
using isochrones,  
%1.030$_{-0.020}^{+0.025}\pm$0.005
1.030 M$_\odot$ adding also lithium,
and  
%1.030$\pm$0.005$\pm$0.005
1.030, including stellar parameters, lithium and the 
mean large frequency separation. The mass derived by \cite{don09}, 1.01$\pm$0.01 M$_\odot$,
is also in agreement with our results within the error bars,
as well as with the mass derived using asteroseismology, 1.02$\pm$0.03 M$_\odot$ \citep{baz11}
and 1.01$\pm$0.03 M$_\odot$ \citep{baz12}. We are performing a more detailed 
seismic analysis of 18 Sco including also new HARPS observations (Bazot et al., in preparation).

The Li abundance (A$_{\rm Li}^{\rm NLTE}$ = 1.62 $\pm$ 0.02 dex) was derived using 
the line list presented in \cite{mel12} and NLTE corrections by \cite{lin09}, 
and is identical to that obtained by \cite{mon13}, 
as the stellar parameters are essentially the same, except for a 1 K difference in 
the effective temperature. We refer the reader to \cite{mon13} for further details,
but we stress here that our Li abundance fits well the trend of Li depletion with
age of several non-standard solar models \citep[e.g.,][]{ct05,don09,xd09,den10}.

\section{Companions around 18 Sco}

18 Sco is included in our HARPS Large Program to search for planets around solar twins,
hence we can evaluate whether planets or a binary component are present.

Radial velocities were obtained with the HARPS instrument and binned to yield one RV value per 
night for a total of 59 nights spanning from May 2004 to February 2014.  
The observations include 20 nights of high-cadence asteroseismic observations 
without simultaneous reference in addition to 39 nights of observations with simultaneous ThAr
reference for instrumental drift correction.  The asteroseismic data have a scatter of 1 m s$^{-1}$ 
throughout the course of a single night.  When a moving average is applied to smooth out 
random noise, a coherent and repeated nightly pattern with an amplitude on order of 2 m s$^{-1}$ emerges.
We conclude that this coherent noise may be instrumental in origin and minimize its effect 
on the data by using only a single data point from each night obtained by a weighted average 
of points from the 4 hours with lowest airmass. 
Additionally, the radial velocities show drifts throughout the course of the two asteroseismic 
observing runs which may be instrumental or, in the case of the May 2012 observations, 
may be a signal from starspots. We make no attempt to remove these drifts due to the uncertainty of their origin.

Activity indices were also calculated for each HARPS spectrum from the \ion{Ca}{2} H \& K lines.  
The activity cycle of 18 Sco is present in the data, with radial velocities increasing at 
times of high activity as photospheric convection is suppressed.  This variation occurs 
on a timescale of 7.6 years in the data, consistent with the previously measured period 
of seven years from photometry and chromospheric activity \citep{hal07}.  
We remove the effect of the activity cycle on the RVs 
by fitting and subtracting a linear correlation between radial velocity and 
continuum-normalized \ion{Ca}{2} H \& K flux ($S_{HK}$).

\begin{figure}
\epsscale{1.1}
\plotone{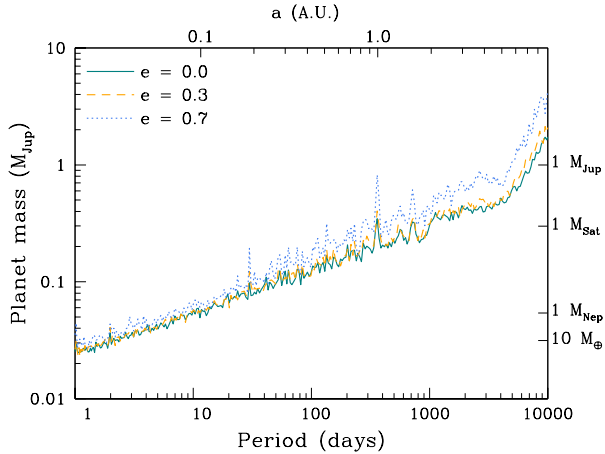}
\caption{Detection limits based on our HARPS data are shown for different
eccentricities (e) as a function of orbital period. Planets above the curves are ruled out.
\label{18Sco_limits}}
\end{figure}

The resulting radial velocity measurements were searched for planet signals with no significant detections.  
We quantify our upper limits on potential planets as follows.  We make a flat-line fit to the data and subtract the offset.  We then fold all 
jitter in the radial velocities into the uncertainties on the residuals by scaling them with the reduced 
chi-squared of the fit.  The residuals are resampled randomly with replacement, and a Keplerian signal of 
fixed planet period and mass is added.  If the resulting simulated RVs are inconsistent with a 
flat-line fit by three or more sigma for 99\% of randomized trials at a given planet 
period and mass, we consider the planet to be excluded by the data.  The resulting exclusion 
limits rule out sub-Neptune-mass objects out to a period of 7 days, and Jupiter-mass objects 
out to approximately 13 years  (Fig. \ref{18Sco_limits}).  As with all ground-based radial 
velocity observations, these data show aliases on timescales around 1 day (due to observing 
nightly), and at approximately 30 and 365 days due to the effects of the lunar cycle and 
seasonal observability on sampling. 

Our HARPS RV measurements also exclude the presence of a binary component, which is 
consistent with no detection of companions around 18 Sco by high-contrast AO imaging \citep{tan10}.

\section{The complex abundance pattern of 18 Sco}
\begin{figure}
\epsscale{1.1}
\plotone{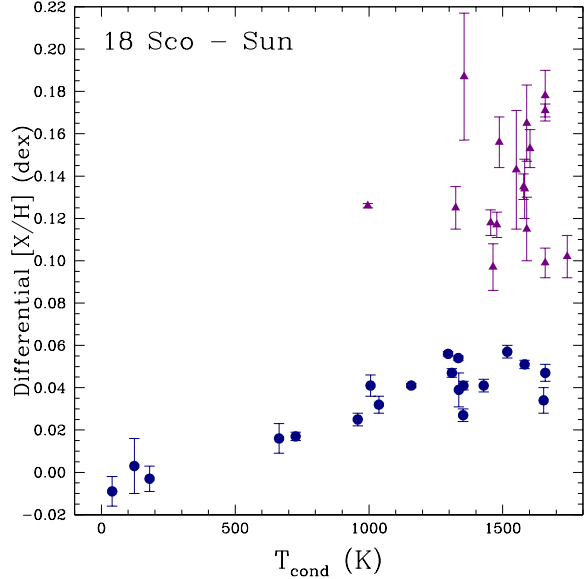}
\caption{Differential abundances of 18 Sco relative to the Sun. Elements with [X/H] $\leq$ 0.06 
(filled circles) have Z $\leq$ 30, i.e, Zinc and lighter elements, while the elements with 
[X/H] $>$ 0.09 (filled triangles) have Z $>$ 30.
\label{18sco1}}
\end{figure}

As can be seen in Fig. \ref{18sco1}, 18 Sco presents a complex abundance pattern. 
On top of the typical trend with condensation temperature seen in other solar twins 
\citep{mel09,ram09a,ram10}, corresponding in Fig. \ref{18sco1} to the group of elements
with enhancements [X/H] $\leq$ 0.06 dex (filled circles), there is a group 
of elements with much larger enhancements (0.09 $<$ [X/H] $<$ 0.19 dex; filled triangles). 
All elements in the latter group are neutron-capture elements.

\begin{figure}
\epsscale{1.1}
\plotone{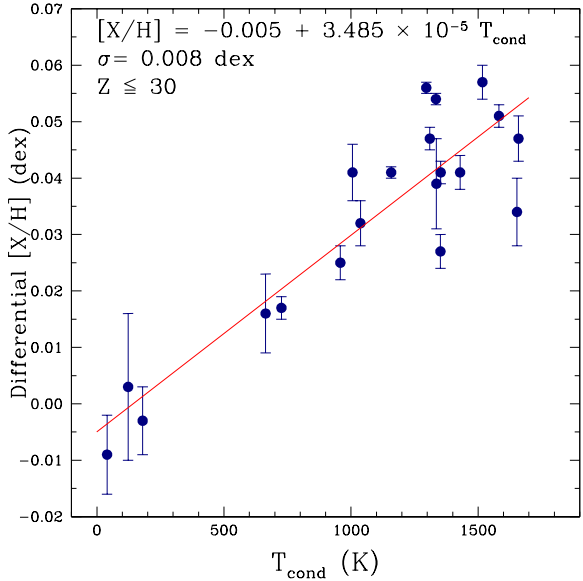}
\caption{Fit of the trend with condensation temperature for elements with Z $\leq$ 30.
The slope has a significance of 9 $\sigma$ and the element-to-element scatter from the fit is only 0.008 dex.
\label{18sco2}}
\end{figure}

In order to understand the large enhancement of the n-capture elements in 18 Sco, we first need to
subtract the trend with condensation temperature, as it is probably related
to the deficiency of refractory elements in the Sun \citep{mel09}; besides, the yields of AGB stars and
SN do not produce such a trend \citep{mel12}. We fit [X/H] vs. condensation temperature 
\citep{lod03} for the lighter elements with Z $\leq$ 30, 
because they only seem affected by the condensation temperature.
The fit (Fig. \ref{18sco2}) results in:

\begin{equation}
 [X/H](Z \leq 30) = -0.005 + 3.485 \times 10^{-5} T_{\rm cond},
\end{equation}

with an element-to-element scatter of only 0.008 dex, which is similar
to the mean total error (0.007 dex) of our differential abundances for
elements with Z $\leq$ 30 (Table \ref{abun}),
showing thus that our small error bars are realistic. 
The significance of the slope is 9 $\sigma$.

\begin{figure}
\epsscale{1.1}
\plotone{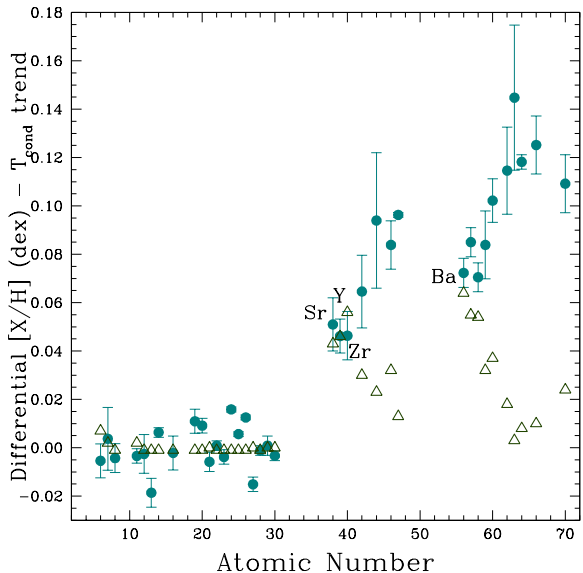}
\caption{The filled circles are the [X/H] ratios in 18 Sco after they have been subtracted from
the condensation temperature trend shown in Fig. \ref{18sco2}. Elements with Z $\leq$ 30 now have  abundance
ratios close to zero but the neutron-capture elements remain enhanced.
The open triangles represent the effect of pollution by an AGB star (see the text for details). 
Although a good match can be
seen for the s-process elements Sr, Y, Zr and Ba, there is a disagreement for other elements.
\label{18sco3}}
\end{figure}

Next, we subtract the above trend from the [X/H] abundances:

\begin{equation}
 [X/H]_T = [X/H] - (-0.005 + 3.485 \times 10^{-5} T_{\rm cond}).
\end{equation}

The [X/H]$_T$ ratios are given in Table \ref{ncapture} and
are shown by filled circles in Fig. \ref{18sco3}. 

We first verify if the observed enhancement in the n-capture elements is due to 
pollution by AGB stars. 
We used  a model of a 3 M$_{\odot}$ AGB star of Z = 0.01 
\citep{kar10}\footnote{Similar models are considered to derive the $s$-process
contribution in the solar system \citep[e.g.,][]{arl99}.}
and diluted the yields of a small fraction of 
AGB ejecta ($\leq$ 1\% of material injected) into a 1 M$_{\odot}$ proto cloud 
of solar composition \citep{asp09}. Then, we computed the enhancement in the abundance ratios
relative to the initial composition of the proto cloud, [X/H]$_{\rm AGB}$.
A dilution of 0.23\% mass of AGB material matched the observed enhancement
in the light $s$-process element\footnote{As common in the literature,
we use the terms $s$-process and $r$-process elements, but rigorously speaking
that is incorrect, as the $s$ and $r$ neutron capture processes are responsible 
for the synthesis of isotopes. The $s$-process and $r$-process elements are deduced 
from decomposition of Solar system material.} 
yttrium. The [X/H]$_{\rm AGB}$ ratios are given in Table \ref{ncapture} and
shown by open triangles in Fig. \ref{18sco3}.
As can be seen, a good match cannot be achieved for all the n-capture elements, 
showing that there is an additional source for the abundance enhancement.
Nevertheless, other $s$-process elements besides Y, such as Sr, Zr and Ba are well fitted, thus, 
the observed enhancement could be due, at least partly, to AGB stars.
In order to find out if the residual enhancement is due to the $r$-process,
we estimated its enrichment in 18 Sco by subtracting the AGB contribution

\begin{equation}
 [X/H]_r = [X/H]_T - [X/H]_{\rm AGB},
\end{equation}

and compare these results with the predicted enhancement based on the
$r$-process fractions in the solar system, $r_{SS} = 1 - s_{SS}$, using the
$s$-fractions ($s_{SS}$) by \cite{sim04} and Bisterzo et al. (2011, updated for a few elements 
by Bisterzo et al. 2013).
Then, we can estimate the $r$-process contribution [X/H]$_r^{SS}$ from

\begin{equation}
 1 - s_{SS} = 10^{[X/H]_r^{SS}/\Delta_T}/10^{[X/H]_T/\Delta_T},
\end{equation}

where we define $\Delta_T$ as the average of the three most enhanced $s$-process and $r$-process elements,
corresponding to $\Delta_T$ = 0.093 dex for the observed [X/H]$_T$ enrichment in 18 Sco (Table \ref{ncapture}).

Therefore, the predicted $r$-process contribution based on the solar system $r$-fractions, [X/H]$_r^{SS}$, is:

\begin{equation}
 [X/H]_r^{SS} = \Delta_T \times \log_{10} (1 - s_{SS}) + [X/H]_T.
\end{equation}

\begin{figure}
\epsscale{1.1}
\plotone{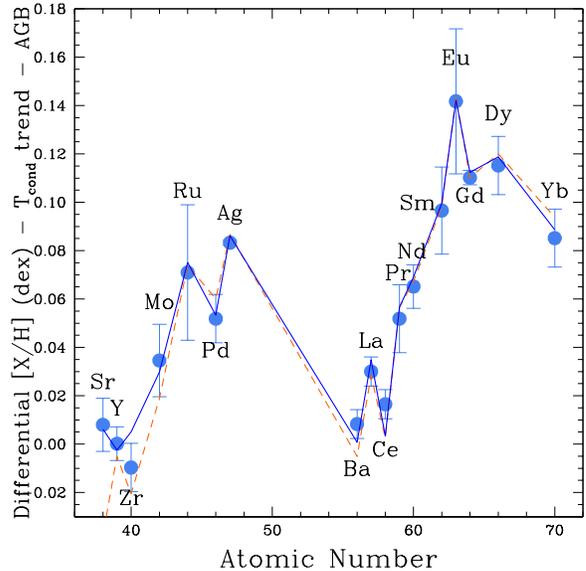}
\caption{The filled circles represent the [X/H] ratios in 18 Sco after they have been subtracted from
the condensation temperature trend (Fig. \ref{18sco2}) and from the AGB contribution (Fig. \ref{18sco3}).
The residual enhancement, [X/H]$_r$ (filled circles), is in extraordinary agreement
with the predicted $r$-process enhancement based on the solar system $r$-process fractions
by \cite{sim04} and \cite{bis11,bis13}, represented by dashed and solid lines, respectively.
\label{18sco3r}}
\end{figure}

\begin{figure}
\epsscale{1.1}
\plotone{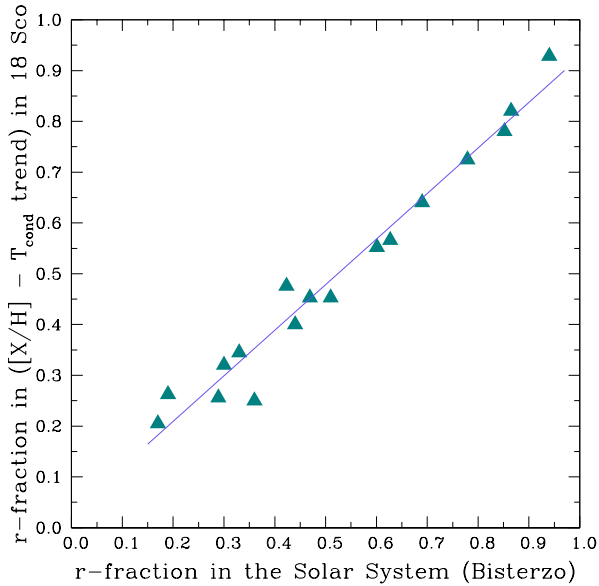}
\caption{The $r$-process fractions in the [X/H]$_T$ abundance ratios in 18 Sco versus
the $r$-fractions in the solar system \citep{bis11,bis13}. The line shows a linear fit
with slope of 0.90 and element-to-element scatter of only 0.04.
\label{18sco3rf}}
\end{figure}

In Fig. \ref{18sco3r}, we compare the ``observed'' $r$-process enhancement [X/H]$_r$ (eq. 7)
with the predicted one based on the solar system $r$-process fractions, [X/H]$_r^{SS}$ (eq. 9).
There is an astonishing agreement, strongly suggesting
that the remaining enhancement is indeed due to the $r$-process.
The impressive agreement, even at the scale of the small fluctuations ($\sim$0.02 dex) among 
nearby n-capture elements (Fig \ref{18sco3r}), also suggests that even for heavy elements 
we suceeded in achieving abundances with errors of about 0.01 dex.

Another way to show that the residual enhancement (after subtracting the AGB contribution) is
due to the $r$-process, is by comparing the $r$-fractions of the [X/H]$_T$ ratios with the
$r$-fractions in the solar system. Since the \cite{bis11,bis13} $s_{SS}$-fractions fit 
somewhat better our [X/H]$_r$ ratios (Fig. \ref{18sco3r}), we will use their values in the comparison.
The ``observed'' $r$-fractions in the [X/H]$_T$ ratios in 18 Sco, are estimated by

\begin{equation}
 r = 10^{[X/H]_r/\Delta_T}/10^{[X/H]_T/\Delta_T}.
\end{equation}

The ``observed'' $r$-fractions and the solar system $r_{SS}$-fractions \citep{bis11,bis13},
are compared in Fig. \ref{18sco3rf}. Again, the agreement is remarkable (slope of 0.90
and element-to-element scattter of only 0.04),
showing that after both the T$_{\rm cond}$ trend and AGB contribution are subtracted,
the remaining material can be explained by the $r$-process.
Our high precision abundances provide independent evidence of the universality of the $r$-process,
i.e., to the remarkable similarity of the relative abundance pattern among different $r$-process elements.
Our finding is similar to what is found in metal-poor stars, where the scaled solar $r$-process pattern 
match the abundances of most neutron-capture elements
\citep[e.g.,][]{cow02,hil02,fre07,sne08,siq13}.

The complex abundance pattern of 18 Sco can be thus explained by the condensation temperature trend, 
the $s$-process and the $r$-process. The excess of refractory elements relative to the Sun seems to be typical of 
solar like stars \citep{mel09,ram09a,ram10,sch11,liu14,mac14}, so 18 Sco looks normal in this regard. 
As we will see below, the additional enhancement in the n-capture elements may be explained by the younger age 
of 18 Sco (Section 4).

A deficiency in the abundances of the $s$-process elements have been reported in our analysis of the pair 
of almost solar twins 16 Cyg \citep{ram11} and the solar twin HIP 102152 \citep{mon13}, which are both 
older than the Sun by 2.5 and 3.7 Gyr, respectively. On the contrary, the s-process elements 
seem enhanced in the young solar twin 18 Sco. Thus, the increasing enhancement of the $s$-process elements 
with decreasing age is probably due to the pollution of successive generations of AGB stars, 
that can be tracked using solar twins spanning ages from 2.9 Gyr (18 Sco) up to 8.2 Gyr (HIP 102152). 
Similar enhancements of the s-process elements have been found in open clusters 
\citep[e.g.,][]{dor09,mai11,yon12,jac13,mis13b}, with Ba showing the clearest trend 
of increasing abundance for younger ages. A similar trend was observed in field stars
\citep{edv93,ben07,das12}, but \cite{mis13} found no dependence of [Ba/Fe] and [La/Fe] with age.
Albeit there is no consensus yet, most evidences 
point out for enhanced $s$-process abundances for decreasing ages.

Regarding the r-process elements, \cite{das12} found a
decrease of [Sm/Fe] for decreasing ages in stars younger than the Sun.
Using both Eu abundances (4129 \AA\ line) and stellar ages determined by \cite{ben05} in thin disk stars,
we found that for ages lower than 9 Gyr there is a flat [Eu/Fe] ratio, i.e., no dependence with stellar age.
Unfortunately, neither \cite{del05} nor \cite{mis13} studied the dependence of their 
[Eu/Fe] ratios with age in thin disk stars. 
Our analysis of two old solar twins \citep{ram11,mon13}, found normal
$r$-process abundances for 16 Cyg B and HIP 102152.
Based on the limited evidence, $r$-process elements 
seem to have a flat behavior with age.

Since 18 Sco is considerably younger than the Sun, the enhancement in the 
$s$-process elements could be due just to normal Galactic chemical evolution.
However, the enhancement of the $r$-process elements
in 18 Sco is more difficult to understand, as, based on our discussion above,
those elements are not expected to be enriched.

The unexpected enhancement of the $r$-process elements in 18 Sco could be attributed to a somewhat higher
contribution of $r$-process ejecta to the natal cloud of 18 Sco than around other solar-like stars.

Our precise abundances show that whatever sources that produced the enrichment in the
n-capture elements, did not produce substantial quantities of elements lighter than Z = 30.
The amount of mass with Z $>$ 30 that would be needed to produced the observed enhancements
is very small, only 2.7 $\times 10^{-8}$ M$_\odot$  and 2.4 $\times 10^{-7}$ M$_\odot$ 
for the $s$- and $r$-process, respectively.

\section{Implications for Chemical Tagging}
Studying the detailed abundance pattern of solar type stars we may be able 
to reconstruct the build up of the Galaxy using chemical tagging \citep{fre02}. 
One of the main problems when tagging individual stars to a given natal cloud may be 
radial migration \citep{sel02}, yet the chemical abundances may be preserved, hence using 
chemical tagging stellar groups or clusters could be reconstructed. 

Based on our precise abundances in 18 Sco, it is clear that different elements should be 
targeted for chemical tagging. First, as many of the highly volatile (C, N, O) and highly 
refractory elements (Al, Ti, Sc, Ca, V, Fe) should be analyzed to determine if there is any 
trend with condensation temperature. If possible it would be important to cover also some 
semi-volatiles (e.g., S, Zn, Na, Cu, K, Mn, P) and some medium refractories 
(e.g., Si, Mg, Cr, Ni, Co, V), to verify if there is a break in the trend with condensation 
temperature \citep{mel09}. Also, it should be important not to discard stars from a given
group due to small discrepant abundances, as those anomalies could be due to either the formation of
terrestrial \citep{mel09} or giant \citep{tuc14,ram11} planets.

Some of the above elements are of course relevant to different nucleosynthetic processes such 
as production of $\alpha$ elements (O, S, Ca, Si, Ti) by SNe II or signatures of 
proton-burning (Na, Mg, Al, O). Li is important as a potential chronometer \citep{bau10,mon13,mel14} 
and to study the transport mechanisms inside stars \citep[e.g.][]{ct05,don09}. 
The study of 18 Sco also shows the importance of including the heavy elements for chemical tagging. 
Ideally, at least some elements between Sr, Zr, Y or Ba should be included to verify the $s$-process 
and some elements among Ru, Pd, Ag, Sm, Gd, Eu or Dy could be explored for the $r$-process.

\section{Conclusions}
We have performed the most precise and complete abundance analysis of 18 Sco. 
Being the brightest of the solar twins, very high S/N, high resolution spectra were 
gathered for 18 Sco and the Sun and a strictly differential line-by-line analysis was performed 
allowing us to achieve a precision of about 0.005 dex in differential abundances.
Additionally, highly precise stellar parameters were obtained, which would be important for 
further modeling of this solar twin using different techniques.
Precise radial velocities were gathered with HARPS, but no planet has
been detected yet.

The complex abundance pattern of 18 Sco shows enhancements (relative to the Sun) 
in the refractory, $s$-process and $r$-process elements. 
After subtracting the trend with condensation temperature
and the contribution from AGB stars, the remaining enhancement shows the same
pattern as the $r$-process in the solar system.
This shows the universality of the $r$-process.
The different contributions to the abundance enrichment in 18 Sco
could be disentangled thanks to the exquisite precision achieved in our work.

18 Sco serves as a testbed for studies of chemical tagging in large samples of stars of 
upcoming surveys, such as HERMES\footnote{http://www.aao.gov.au/HERMES/},
which plans to survey about a million stars at high spectral resolution.

\acknowledgments

J.M. and T.R.M. acknowledge support from FAPESP (2012/24392-2 and 2010/19810-4).
M.B. and J.B. acknowledge support for this work from the NSF (grant
number AST-1313119)

%{\it Facilities:} \facility{VLT (UVES)}, \facility{Keck (HIRES)}, \facility{ESO:3.6m (HARPS)}.

%\clearpage

%% the file, uncomment the \plotone, \plottwo, and
%% If you need a layout that cannot be achieved with \plotone or
%% \plottwo, you can invoke the graphicx package directly with the
%% \includegraphics command or use \plotfiddle. For more information,
%% please see the tutorial on "Using Electronic Art with AASTeX" in the
%% documentation section at the AASTeX Web site,
%% http://www.journals.uchicago.edu/AAS/AASTeX.
%%
%% The examples below also include sample markup for submission of
%% supplemental electronic materials. As always, be sure to check
%% the instructions to authors for the journal you are submitting to
%% for specific submissions guidelines as they vary from

%% This example uses \plotone to include an EPS file scaled to
%% 80% of its natural size with \epsscale. Its caption
%% has been written to indicate that additional figure parts will be
%% available in the electronic journal.

\begin{longtable}{ccccccc}
\caption{Adopted atomic data and equivalent widths} \tabularnewline
\label{EW}
\centering 
%\hline 
{Wavelength}  & ion & $\chi_{exc}$ & log $gf$ & $C_6$ & EW & EW  \\
\hline
 (\AA)         &     & (eV)         &          &       & 18 Sco & Sun \\
\hline
\endfirsthead
\caption{Continued.} \\
\hline 
{Wavelength} & ion & $\chi_{exc}$ & log $gf$ & $C_6$ & EW & EW   \\
\hline
 (\AA) &     & (eV) &     &   & 18 Sco & Sun  \\
\hline
\endhead
\hline
\endfoot
\hline
\endlastfoot
 5044.211 & 26.0 & 2.8512 &  -2.058 &  0.271E-30  &  74.8  &  74.3 \\
 5054.642 & 26.0 & 3.640  &  -1.921 &  0.468E-31  &  40.9  &  40.5 \\
 5127.359 & 26.0 & 0.915  &  -3.307 &  0.184E-31  &  97.5  &  96.1 \\
 5127.679 & 26.0 & 0.052  &  -6.125 &  0.12E-31   &  18.9  &  19.1 \\
 5198.711 & 26.0 & 2.223  &  -2.135 &  0.461E-31  &  99.3  &  98.1 \\
 5225.525 & 26.0 & 0.1101 &  -4.789 &  0.123E-31  &  72.5  &  72.1 \\
 5242.491 & 26.0 & 3.634  &  -0.967 &  0.495E-31  &  88.3  &  86.9 \\
 5247.050 & 26.0 & 0.0872 &  -4.946 &  0.122E-31  &  67.4  &  66.9 \\
 5250.208 & 26.0 & 0.1212 &  -4.938 &  0.123E-31  &  66.3  &  65.9 \\
 5295.312 & 26.0 & 4.415  &  -1.49  &  0.654E-30  &  31.0  &  30.3 \\
 5322.041 & 26.0 & 2.279  &  -2.80  &  0.429E-31  &  62.9  &  61.5 \\
 5373.709 & 26.0 & 4.473  &  -0.77  &  0.704E-30  &  65.2  &  63.9 \\
 5379.574 & 26.0 & 3.694  &  -1.514 &  0.502E-31  &  62.9  &  61.5 \\
 5386.334 & 26.0 & 4.154  &  -1.74  &  0.527E-30  &  34.8  &  33.6 \\
 5466.396 & 26.0 & 4.371  &  -0.565 &  0.440E-30  &  81.3  &  79.4 \\
 5466.987 & 26.0 & 3.573  &  -2.23  &    2.8      &  35.8  &  35.2 \\
 5522.446 & 26.0 & 4.209  &  -1.31  &  0.302E-30  &  45.8  &  43.7 \\
 5546.506 & 26.0 & 4.371  &  -1.18  &  0.391E-30  &  52.6  &  51.4 \\
 5560.211 & 26.0 & 4.434  &  -1.16  &  0.479E-30  &  53.8  &  52.0 \\
 5577.02  & 26.0 & 5.0331 &  -1.455 &    2.8      &  11.9  &  11.2 \\
 5618.633 & 26.0 & 4.209  &  -1.276 &  0.290E-30  &  51.8  &  50.2 \\
 5636.696 & 26.0 & 3.640  &  -2.56  &  0.519E-31  &  20.7  &  19.7 \\
 5638.262 & 26.0 & 4.220  &  -0.81  &  0.288E-30  &  78.8  &  77.6 \\
 5649.987 & 26.0 & 5.0995 &  -0.8   &  0.277E-30  &  37.8  &  35.9 \\
 5651.469 & 26.0 & 4.473  &  -1.75  &  0.483E-30  &  20.3  &  18.9 \\
 5661.348 & 26.0 & 4.2843 &  -1.756 &  0.324E-30  &  24.4  &  23.0 \\
 5679.023 & 26.0 & 4.652  &  -0.75  &  0.813E-30  &  61.7  &  59.6 \\
 5696.089 & 26.0 & 4.548  &  -1.720 &  0.578E-30  &  14.5  &  13.7 \\
 5701.544 & 26.0 & 2.559  &  -2.216 &  0.495E-31  &  86.4  &  84.7 \\
 5705.464 & 26.0 & 4.301  &  -1.355 &  0.302E-30  &  39.7  &  38.1 \\
 5855.076 & 26.0 & 4.6075 &  -1.478 &  0.574E-30  &  23.8  &  22.9 \\
 5905.672 & 26.0 & 4.652  &  -0.69  &  0.623E-30  &  61.7  &  60.1 \\
 5916.247 & 26.0 & 2.453  &  -2.936 &  0.429E-31  &  55.4  &  55.5 \\
 5927.789 & 26.0 & 4.652  &  -1.04  &  0.607E-30  &  46.0  &  43.6 \\
 5934.655 & 26.0 & 3.928  &  -1.07  &  0.569E-30  &  79.1  &  77.9 \\
 5956.694 & 26.0 & 0.8589 &  -4.605 &  0.155E-31  &  52.3  &  52.8 \\
 5987.065 & 26.0 & 4.795  &  -0.212 &  0.155E-31  &  70.3  &  68.2 \\
 6003.012 & 26.0 & 3.881  &  -1.060 &  0.483E-30  &  85.5  &  84.5 \\
 6005.541 & 26.0 & 2.588  &  -3.43  &    2.8      &  23.9  &  23.0 \\
 6024.058 & 26.0 & 4.548  &  -0.02  &  0.388E-30  & 114.0  & 111.2 \\
 6027.050 & 26.0 & 4.0758 &  -1.09  &    2.8      &  65.7  &  64.0 \\
 6056.005 & 26.0 & 4.733  &  -0.45  &  0.679E-30  &  74.2  &  72.3 \\
 6065.482 & 26.0 & 2.6085 &  -1.530 &  0.471E-31  & 118.8  & 117.3 \\
 6079.009 & 26.0 & 4.652  &  -1.10  &  0.513E-30  &  47.5  &  46.5 \\
 6082.711 & 26.0 & 2.223  &  -3.573 &  0.327E-31  &  35.3  &  34.0 \\
 6093.644 & 26.0 & 4.607  &  -1.30  &  0.441E-30  &  32.8  &  30.8 \\
 6096.665 & 26.0 & 3.9841 &  -1.81  &  0.575E-30  &  39.3  &  37.6 \\
 6151.618 & 26.0 & 2.1759 &  -3.299 &  0.255E-31  &  51.2  &  49.9 \\
 6157.728 & 26.0 & 4.076  &  -1.22  &    2.8      &  64.7  &  62.7 \\
 6165.360 & 26.0 & 4.1426 &  -1.46  &    2.8      &  47.4  &  45.4 \\
 6173.335 & 26.0 & 2.223  &  -2.880 &  0.265E-31  &  69.0  &  68.7 \\
 6187.990 & 26.0 & 3.943  &  -1.67  &  0.490E-30  &  49.4  &  47.6 \\
 6200.313 & 26.0 & 2.6085 &  -2.437 &  0.458E-31  &  74.6  &  73.6 \\
 6213.430 & 26.0 & 2.2227 &  -2.52  &  0.262E-31  &  83.9  &  82.7 \\
 6219.281 & 26.0 & 2.198  &  -2.433 &  0.258E-31  &  91.2  &  90.3 \\
 6226.736 & 26.0 & 3.883  &  -2.1   &  0.415E-30  &  31.5  &  30.1 \\
 6240.646 & 26.0 & 2.2227 &  -3.233 &  0.314E-31  &  49.3  &  48.7 \\
 6252.555 & 26.0 & 2.4040 &  -1.687 &  0.384E-31  & 123.1  & 121.6 \\
 6265.134 & 26.0 & 2.1759 &  -2.550 &  0.248E-31  &  86.9  &  85.9 \\
 6270.225 & 26.0 & 2.8580 &  -2.54  &  0.458E-31  &  52.9  &  52.2 \\
 6271.279 & 26.0 & 3.332  &  -2.703 &  0.278E-30  &  25.3  &  24.6 \\
 6380.743 & 26.0 & 4.186  &  -1.376 &    2.8      &  54.3  &  53.2 \\
 6392.539 & 26.0 & 2.279  &  -4.03  &  0.338E-31  &  17.3  &  16.7 \\
 6430.846 & 26.0 & 2.1759 &  -2.006 &  0.242E-31  & 113.4  & 114.0 \\
 6498.939 & 26.0 & 0.9581 &  -4.699 &  0.153E-31  &  47.2  &  46.5 \\
 6593.871 & 26.0 & 2.4326 &  -2.422 &  0.369E-31  &  86.6  &  86.3 \\
 6597.561 & 26.0 & 4.795  &  -0.98  &  0.476E-30  &  46.2  &  44.7 \\
 6625.022 & 26.0 & 1.011  &  -5.336 &    2.8      &  15.8  &  14.9 \\
 6677.987 & 26.0 & 2.692  &  -1.418 &  0.346E-31  & 132.8  & 129.9 \\
 6703.567 & 26.0 & 2.7585 &  -3.023 &  0.366E-31  &  38.3  &  37.6 \\
 6705.102 & 26.0 & 4.607  &  -0.98  &    2.8      &  47.9  &  46.3 \\
 6710.319 & 26.0 & 1.485  &  -4.88  &  0.201E-31  &  15.7  &  15.7 \\
 6713.745 & 26.0 & 4.795  &  -1.40  &  0.430E-30  &  22.3  &  21.2 \\
 6725.357 & 26.0 & 4.103  &  -2.19  &  0.482E-30  &  19.2  &  18.6 \\
 6726.667 & 26.0 & 4.607  &  -1.03  &  0.482E-30  &  48.4  &  47.4 \\
 6733.151 & 26.0 & 4.638  &  -1.47  &  0.341E-30  &  28.7  &  27.0 \\
 6739.522 & 26.0 & 1.557  &  -4.79  &  0.210E-31  &  13.2  &  12.4 \\
 6750.152 & 26.0 & 2.4241 &  -2.621 &  0.411E-31  &  75.3  &  75.3 \\
 6752.707 & 26.0 & 4.638  &  -1.204 &  0.337E-30  &  39.0  &  36.8 \\
 6793.259 & 26.0 & 4.076  &  -2.326 &    2.8      &  14.3  &  13.0 \\
 6806.845 & 26.0 & 2.727  &  -3.11  &  0.346E-31  &  35.8  &  35.4 \\
 6810.263 & 26.0 & 4.607  &  -0.986 &  0.450E-30  &  53.1  &  50.6 \\
 6837.006 & 26.0 & 4.593  &  -1.687 &  0.246E-31  &  20.0  &  18.9 \\
 6839.830 & 26.0 & 2.559  &  -3.35  &  0.395E-31  &  32.5  &  31.6 \\
 6843.656 & 26.0 & 4.548  &  -0.86  &  0.294E-30  &  64.1  &  63.3 \\
 6858.150 & 26.0 & 4.607  &  -0.930 &  0.324E-30  &  53.6  &  52.4 \\
 5197.577 & 26.1 & 3.2306 &  -2.22  &  0.869E-32  &  84.0  &  80.7 \\
 5234.625 & 26.1 & 3.2215 &  -2.18  &  0.869E-32  &  83.5  &  80.7 \\
 5264.812 & 26.1 & 3.2304 &  -3.13  &  0.943E-32  &  48.6  &  45.8 \\
 5325.553 & 26.1 & 3.2215 &  -3.16  &  0.857E-32  &  42.7  &  41.3 \\
 5414.073 & 26.1 & 3.2215 &  -3.58  &  0.930E-32  &  28.5  &  26.7 \\
 5425.257 & 26.1 & 3.1996 &  -3.22  &  0.845E-32  &  43.0  &  41.3 \\
 6084.111 & 26.1 & 3.1996 &  -3.79  &  0.787E-32  &  22.6  &  20.9 \\
 6247.557 & 26.1 & 3.8918 &  -2.30  &  0.943E-32  &  54.6  &  52.9 \\
 6369.462 & 26.1 & 2.8912 &  -4.11  &  0.742E-32  &  20.4  &  18.7 \\
 6416.919 & 26.1 & 3.8918 &  -2.64  &  0.930E-32  &  42.6  &  40.5 \\
 6432.680 & 26.1 & 2.8912 &  -3.57  &  0.742E-32  &  43.9  &  42.4 \\
 6456.383 & 26.1 & 3.9036 &  -2.05  &  0.930E-32  &  64.9  &  62.4 \\
 5052.167 & 06.0 & 7.685  &  -1.24  &    2.8      &  34.0  &  34.5 \\
 5380.337 & 06.0 & 7.685  &  -1.57  &    2.8      &  20.5  &  20.6 \\
 6587.61  & 06.0 & 8.537  &  -1.05  &    2.8      &  15.6  &  15.2 \\
 7111.47  & 06.0 & 8.640  &  -1.07  &  0.291E-29  &  14.9  &  14.3 \\
 7113.179 & 06.0 & 8.647  &  -0.76  &  0.297E-29  &  24.9  &  23.6 \\
 7771.944 & 08.0 & 9.146  &   0.37  &  0.841E-31  &  73.0  &  71.1 \\
 7774.166 & 08.0 & 9.146  &   0.22  &  0.841E-31  &  65.6  &  63.7 \\
 7775.388 & 08.0 & 9.146  &   0.00  &  0.841E-31  &  51.7  &  51.1 \\
 4751.822 & 11.0 & 2.1044 &  -2.078 &    2.8      &  11.9  &  11.6 \\
 5148.838 & 11.0 & 2.1023 &  -2.044 &    2.8      &  11.8  &  11.7 \\
 6154.225 & 11.0 & 2.1023 &  -1.547 &    2.8      &  36.9  &  36.9 \\
 6160.747 & 11.0 & 2.1044 &  -1.246 &    2.8      &  54.0  &  54.3 \\
 4571.095 & 12.0 & 0.000  &  -5.623 &    2.8      & 105.6  & 106.0 \\
 4730.040 & 12.0 & 4.340  &  -2.389 &    2.8      &  69.3  &  69.0 \\
 5711.088 & 12.0 & 4.345  &  -1.729 &    2.8      & 106.4  & 105.6 \\
 6319.236 & 12.0 & 5.108  &  -2.165 &    2.8      &  27.6  &  25.6 \\
 6696.018 & 13.0 & 3.143  &  -1.481 &    2.8      &  37.0  &  36.3 \\
 6698.667 & 13.0 & 3.143  &  -1.782 &    2.8      &  21.7  &  21.2 \\
 7835.309 & 13.0 & 4.021  &  -0.68  &    2.8      &  43.6  &  43.6 \\
 7836.134 & 13.0 & 4.021  &  -0.45  &    2.8      &  60.9  &  57.7 \\
 8772.866 & 13.0 & 4.0215 &  -0.38  &  0.971E-29  &  75.0  &  73.9 \\
 8773.896 & 13.0 & 4.0216 &  -0.22  &  0.971E-29  &  95.3  &  92.3 \\
 5488.983 & 14.0 & 5.614  &  -1.69  &    2.8      &  21.4  &  19.8 \\
 5517.540 & 14.0 & 5.080  &  -2.496 &    2.8      &  15.3  &  13.9 \\
 5645.611 & 14.0 & 4.929  &  -2.04  &    2.8      &  37.7  &  36.1 \\
 5665.554 & 14.0 & 4.920  &  -1.94  &    2.8      &  43.2  &  41.1 \\
 5684.484 & 14.0 & 4.953  &  -1.55  &    2.8      &  64.3  &  62.0 \\
 5690.425 & 14.0 & 4.929  &  -1.77  &    2.8      &  51.2  &  49.3 \\
 5701.104 & 14.0 & 4.930  &  -1.95  &    2.8      &  41.4  &  40.5 \\
 5793.073 & 14.0 & 4.929  &  -1.96  &    2.8      &  45.8  &  44.0 \\
 6125.021 & 14.0 & 5.614  &  -1.50  &    2.8      &  33.8  &  32.1 \\
 6145.015 & 14.0 & 5.616  &  -1.41  &    2.8      &  40.8  &  39.1 \\
 6243.823 & 14.0 & 5.616  &  -1.27  &    2.8      &  47.0  &  44.6 \\
 6244.476 & 14.0 & 5.616  &  -1.32  &    2.8      &  48.3  &  45.8 \\
 6721.848 & 14.0 & 5.862  &  -1.12  &    2.8      &  48.0  &  44.8 \\
 6741.63  & 14.0 & 5.984  &  -1.65  &    2.8      &  17.3  &  16.4 \\
 6046.000 & 16.0 & 7.868  &  -0.15  &    2.8      &  22.2  &  20.0 \\
 6052.656 & 16.0 & 7.870  &  -0.4   &    2.8      &  12.8  &  12.3 \\
 6743.54  & 16.0 & 7.866  &  -0.6   &    2.8      &   9.0  &   8.5 \\
 6757.153 & 16.0 & 7.870  &  -0.15  &    2.8      &  20.1  &  19.5 \\
 8693.93  & 16.0 & 7.870  &  -0.44  &  0.151E-29  &  13.2  &  12.7 \\
 8694.62  & 16.0 & 7.870  &   0.1   &  0.151E-29  &  30.3  &  28.6 \\
 7698.974 & 19.0 & 0.000  &  -0.168 &  0.104E-30  & 157.3  & 157.0 \\
 5260.387 & 20.0 & 2.521  &  -1.719 &  0.727E-31  &  35.9  &  34.0 \\
 5512.980 & 20.0 & 2.933  &  -0.464 &    2.8      &  88.1  &  85.7 \\
 5581.965 & 20.0 & 2.5229 &  -0.555 &  0.640E-31  &  96.5  &  94.8 \\
 5590.114 & 20.0 & 2.521  &  -0.571 &  0.636E-31  &  94.7  &  92.4 \\
 5867.562 & 20.0 & 2.933  &  -1.57  &    2.8      &  25.1  &  24.0 \\
 6166.439 & 20.0 & 2.521  &  -1.142 &  0.595E-30  &  71.9  &  70.2 \\
 6169.042 & 20.0 & 2.523  &  -0.797 &  0.595E-30  &  95.4  &  92.7 \\
 6455.598 & 20.0 & 2.523  &  -1.34  &  0.509E-31  &  58.0  &  56.8 \\
 6471.662 & 20.0 & 2.525  &  -0.686 &  0.509E-31  &  94.4  &  93.3 \\
 6499.650 & 20.0 & 2.523  &  -0.818 &  0.505E-31  &  88.6  &  86.5 \\
 4743.821 & 21.0 & 1.4478 &   0.35  &  0.597E-31  &   9.3  &   9.1 \\
 5081.57  & 21.0 & 1.4478 &   0.30  &    2.8      &   9.6  &   9.1 \\
 5520.497 & 21.0 & 1.8649 &   0.55  &    2.8      &   7.4  &   7.4 \\
 5671.821 & 21.0 & 1.4478 &   0.55  &    2.8      &  15.3  &  15.2 \\
 5526.820 & 21.1 & 1.770  &   0.140 &    2.8      &  78.5  &  76.6 \\
 5657.87  & 21.1 & 1.507  &  -0.30  &    2.8      &  70.3  &  68.6 \\
 5684.19  & 21.1 & 1.507  &  -0.95  &    2.8      &  39.4  &  38.6 \\
 6245.63  & 21.1 & 1.507  &  -1.030 &    2.8      &  36.4  &  35.3 \\
 6279.76  & 21.1 & 1.500  &  -1.2   &    2.8      &  30.6  &  30.3 \\
 6320.843 & 21.1 & 1.500  &  -1.85  &    2.8      &   9.7  &   9.0 \\
 6604.578 & 21.1 & 1.3569 &  -1.15  &    2.8      &  39.1  &  37.1 \\
 4465.802 & 22.0 & 1.7393 &  -0.163 &  0.398E-31  &  40.6  &  40.4 \\
 4555.485 & 22.0 & 0.8484 &  -0.488 &  0.442E-31  &  66.6  &  66.1 \\
 4758.120 & 22.0 & 2.2492 &   0.425 &  0.384E-31  &  45.2  &  45.2 \\
 4759.272 & 22.0 & 2.2555 &   0.514 &  0.386E-31  &  47.1  &  46.9 \\
 4820.410 & 22.0 & 1.5024 &  -0.439 &  0.378E-31  &  45.9  &  44.3 \\
 4913.616 & 22.0 & 1.8731 &   0.161 &  0.386E-31  &  54.5  &  52.1 \\
 5022.871 & 22.0 & 0.8258 &  -0.434 &  0.358E-31  &  73.0  &  72.6 \\
 5113.448 & 22.0 & 1.443  &  -0.783 &  0.306E-31  &  28.6  &  27.5 \\
 5147.479 & 22.0 & 0.0000 &  -2.012 &  0.208E-31  &  37.6  &  37.5 \\
 5219.700 & 22.0 & 0.021  &  -2.292 &  0.208E-31  &  29.0  &  29.1 \\
 5295.774 & 22.0 & 1.0665 &  -1.633 &  0.258E-31  &  13.6  &  13.3 \\
 5490.150 & 22.0 & 1.460  &  -0.933 &  0.541E-31  &  22.9  &  22.1 \\
 5739.464 & 22.0 & 2.249  &  -0.60  &  0.386E-31  &   8.7  &   8.5 \\
 5866.452 & 22.0 & 1.066  &  -0.840 &  0.216E-31  &  48.6  &  48.0 \\
 6091.174 & 22.0 & 2.2673 &  -0.423 &  0.389E-31  &  16.3  &  15.8 \\
 6126.217 & 22.0 & 1.066  &  -1.424 &  0.206E-31  &  23.1  &  22.8 \\
 6258.104 & 22.0 & 1.443  &  -0.355 &  0.481E-31  &  52.6  &  52.3 \\
 6261.101 & 22.0 & 1.429  &  -0.479 &  0.468E-31  &  49.8  &  49.1 \\
 4470.857 & 22.1 & 1.1649 &  -2.06  &    2.8      &  64.8  &  64.0 \\
 4544.028 & 22.1 & 1.2429 &  -2.53  &    2.8      &  44.4  &  41.5 \\
 4583.408 & 22.1 & 1.165  &  -2.87  &    2.8      &  33.8  &  32.2 \\
 4636.33  & 22.1 & 1.16   &  -3.152 &    2.8      &  20.8  &  20.3 \\
 4657.212 & 22.1 & 1.243  &  -2.47  &    2.8      &  47.8  &  46.4 \\
 4779.985 & 22.1 & 2.0477 &  -1.26  &    2.8      &  65.8  &  64.9 \\
 4865.611 & 22.1 & 1.116  &  -2.81  &    2.8      &  41.5  &  40.3 \\
 4874.014 & 22.1 & 3.095  &  -0.9   &    2.8      &  37.9  &  36.7 \\
 4911.193 & 22.1 & 3.123  &  -0.537 &    2.8      &  53.8  &  52.7 \\
 5211.54  & 22.1 & 2.59   &  -1.49  &    2.8      &  35.1  &  33.5 \\
 5336.778 & 22.1 & 1.582  &  -1.630 &    2.8      &  73.8  &  72.2 \\
 5381.015 & 22.1 & 1.565  &  -1.97  &    2.8      &  62.2  &  60.1 \\
 5418.767 & 22.1 & 1.582  &  -2.11  &    2.8      &  51.1  &  49.1 \\
 4875.486 & 23.0 & 0.040  &  -0.81  &  0.198E-31  &  46.1  &  46.6 \\
 5670.85  & 23.0 & 1.080  &  -0.42  &  0.358E-31  &  19.6  &  19.7 \\
 5727.046 & 23.0 & 1.081  &  -0.011 &  0.435E-31  &  40.2  &  40.1 \\
 6039.73  & 23.0 & 1.063  &  -0.65  &  0.398E-31  &  13.5  &  13.0 \\
 6081.44  & 23.0 & 1.051  &  -0.578 &  0.389E-31  &  14.2  &  14.4 \\
 6090.21  & 23.0 & 1.080  &  -0.062 &  0.398E-31  &  34.4  &  34.6 \\
 6119.528 & 23.0 & 1.064  &  -0.320 &  0.389E-31  &  21.9  &  21.8 \\
 6199.20  & 23.0 & 0.286  &  -1.28  &  0.196E-31  &  13.7  &  13.8 \\
 6251.82  & 23.0 & 0.286  &  -1.34  &  0.196E-31  &  14.4  &  14.9 \\
 4801.047 & 24.0 & 3.1216 &  -0.130 &  0.452E-31  &  51.0  &  49.3 \\
 4936.335 & 24.0 & 3.1128 &  -0.25  &  0.432E-31  &  47.1  &  45.4 \\
 5214.140 & 24.0 & 3.3694 &  -0.74  &  0.206E-31  &  18.2  &  17.6 \\
 5238.964 & 24.0 & 2.709  &  -1.27  &  0.519E-31  &  17.2  &  16.5 \\
 5247.566 & 24.0 & 0.960  &  -1.59  &  0.392E-31  &  83.7  &  82.4 \\
 5272.007 & 24.0 & 3.449  &  -0.42  &  0.315E-30  &  26.3  &  24.9 \\
 5287.20  & 24.0 & 3.438  &  -0.87  &  0.309E-30  &  11.7  &  10.8 \\
 5296.691 & 24.0 & 0.983  &  -1.36  &  0.392E-31  &  94.2  &  93.6 \\
 5300.744 & 24.0 & 0.982  &  -2.13  &  0.392E-31  &  61.0  &  60.4 \\
 5345.801 & 24.0 & 1.0036 &  -0.95  &  0.392E-31  & 115.2  & 113.3 \\
 5348.312 & 24.0 & 1.0036 &  -1.21  &  0.392E-31  & 101.1  & 100.4 \\
 5783.08  & 24.0 & 3.3230 &  -0.43  &  0.802E-30  &  33.4  &  32.1 \\
 5783.87  & 24.0 & 3.3223 &  -0.295 &  0.798E-30  &  46.3  &  44.9 \\
 6661.08  & 24.0 & 4.1926 &  -0.19  &  0.467E-30  &  13.9  &  13.4 \\
 4588.199 & 24.1 & 4.071  &  -0.594 &    2.8      &  71.5  &  69.3 \\
 4592.049 & 24.1 & 4.073  &  -1.252 &    2.8      &  49.0  &  46.6 \\
 5237.328 & 24.1 & 4.073  &  -1.087 &    2.8      &  54.9  &  52.7 \\
 5246.767 & 24.1 & 3.714  &  -2.436 &    2.8      &  16.8  &  15.4 \\
 5305.870 & 24.1 & 3.827  &  -1.97  &    2.8      &  27.7  &  25.6 \\
 5308.41  & 24.1 & 4.0712 &  -1.846 &    2.8      &  28.7  &  26.8 \\
 5502.067 & 24.1 & 4.168  &  -2.049 &    2.8      &  20.1  &  18.3 \\
 5004.891 & 25.0 & 2.9197 &  -1.63  &  0.314E-31  &  14.3  &  13.9 \\
 5399.470 & 25.0 & 3.85   &  -0.104 &    2.8      &  40.6  &  39.3 \\
 6013.49  & 25.0 & 3.073  &  -0.251 &    2.8      &  87.1  &  86.0 \\
 6016.64  & 25.0 & 3.073  &  -0.084 &    2.8      &  96.7  &  96.3 \\
 6021.79  & 25.0 & 3.076  &  +0.034 &    2.8      &  90.4  &  89.6 \\
 5212.691 & 27.0 & 3.5144 &  -0.11  &  0.339E-30  &  20.6  &  20.9 \\
 5247.911 & 27.0 & 1.785  &  -2.08  &  0.327E-31  &  17.9  &  18.2 \\
 5301.039 & 27.0 & 1.710  &  -1.99  &  0.301E-31  &  19.8  &  20.4 \\
 5342.695 & 27.0 & 4.021  &   0.54  &    2.8      &  33.9  &  33.7 \\
 5483.352 & 27.0 & 1.7104 &  -1.49  &  0.289E-31  &  51.2  &  51.5 \\
 5530.774 & 27.0 & 1.710  &  -2.23  &  0.226E-31  &  19.4  &  20.4 \\
 5647.23  & 27.0 & 2.280  &  -1.56  &  0.414E-31  &  14.2  &  14.3 \\
 6189.00  & 27.0 & 1.710  &  -2.46  &  0.206E-31  &  10.7  &  11.1 \\
 6454.995 & 27.0 & 3.6320 &  -0.25  &  0.378E-30  &  14.0  &  14.4 \\
 4953.208 & 28.0 & 3.7397 &  -0.66  &  0.325E-30  &  57.6  &  57.2 \\
 5010.938 & 28.0 & 3.635  &  -0.87  &  0.390E-30  &  50.7  &  50.1 \\
 5176.560 & 28.0 & 3.8982 &  -0.44  &  0.384E-30  &  59.9  &  59.0 \\
 5589.358 & 28.0 & 3.898  &  -1.14  &  0.398E-30  &  28.9  &  27.9 \\
 5643.078 & 28.0 & 4.164  &  -1.25  &  0.379E-30  &  15.7  &  15.1 \\
 5805.217 & 28.0 & 4.1672 &  -0.64  &  0.410E-30  &  44.3  &  42.6 \\
 6086.282 & 28.0 & 4.266  &  -0.51  &  0.406E-30  &  45.8  &  44.0 \\
 6108.116 & 28.0 & 1.676  &  -2.44  &  0.248E-31  &  65.7  &  65.3 \\
 6130.135 & 28.0 & 4.266  &  -0.96  &  0.391E-30  &  23.8  &  22.7 \\
 6176.811 & 28.0 & 4.088  &  -0.26  &  0.392E-30  &  65.0  &  64.4 \\
 6177.242 & 28.0 & 1.826  &  -3.51  &    2.8      &  14.8  &  15.4 \\
 6204.604 & 28.0 & 4.088  &  -1.14  &  0.277E-30  &  22.7  &  22.5 \\
 6223.984 & 28.0 & 4.105  &  -0.98  &  0.393E-30  &  28.8  &  27.9 \\
 6378.25  & 28.0 & 4.1535 &  -0.90  &  0.391E-30  &  33.2  &  32.1 \\
 6643.630 & 28.0 & 1.676  &  -2.1   &  0.214E-31  &  94.1  &  93.9 \\
 6767.772 & 28.0 & 1.826  &  -2.17  &    2.8      &  80.1  &  79.7 \\
 6772.315 & 28.0 & 3.657  &  -0.99  &  0.356E-30  &  52.0  &  51.2 \\
 7727.624 & 28.0 & 3.678  &  -0.4   &  0.343E-30  &  93.3  &  93.2 \\
 7797.586 & 28.0 & 3.89   &  -0.34  &    2.8      &  80.3  &  78.6 \\
 5105.541 & 29.0 & 1.39   &  -1.516 &    2.8      &  93.1  &  94.1 \\
 5218.197 & 29.0 & 3.816  &   0.476 &    2.8      &  48.8  &  48.4 \\
 5220.066 & 29.0 & 3.816  &  -0.448 &    2.8      &  17.5  &  17.0 \\
 7933.13  & 29.0 & 3.79   &  -0.368 &    2.8      &  28.2  &  28.1 \\
 4722.159 & 30.0 & 4.03   &  -0.38  &    2.8      &  69.9  &  70.0 \\
 4810.534 & 30.0 & 4.08   &  -0.16  &    2.8      &  74.3  &  74.2 \\
 6362.35  & 30.0 & 5.79   &   0.14  &    2.8      &  23.4  &  23.0 \\
 4607.338 & 38.0 & 0.00   &   0.283 &  6.557E-32  &  47.2  &  44.9 \\
 4854.867 & 39.1 & 0.9923 &  -0.38  &    2.8      &  57.7  &  54.4 \\
 4883.685 & 39.1 & 1.0841 &   0.07  &    2.8      &  62.0  &  59.3 \\
 4900.110 & 39.1 & 1.0326 &  -0.09  &    2.8      &  58.7  &  55.9 \\
 5087.420 & 39.1 & 1.0841 &  -0.17  &    2.8      &  51.1  &  47.7 \\
 5200.413 & 39.1 & 0.9923 &  -0.57  &    2.8      &  42.0  &  38.4 \\
 4050.320 & 40.1 & 0.713  &  -1.06  &    2.8      &  26.2  &  23.9 \\
 4208.980 & 40.1 & 0.713  &  -0.51  &    2.8      &  45.5  &  42.1 \\
 4442.992 & 40.1 & 1.486  &  -0.42  &    2.8      &  26.8  &  24.4 \\
 3158.16  & 42.0 & 0.000  &  -0.31  &    2.8      &  13.3  &  11.6 \\
 3193.97  & 42.0 & 0.000  &   0.07  &    2.8      &  16.2  &  15.0 \\
 3436.736 & 44.0 & 0.148  &   0.165 &    2.8      &   8.8  &   7.0 \\
 3498.942 & 44.0 & 0.000  &   0.322 &    2.8      &  27.4  &  25.4 \\
 3242.698 & 46.0 & 0.8138 &   0.07  &    2.8      &  27.2  &  25.5 \\
 3404.576 & 46.0 & 0.8138 &   0.33  &    2.8      &  32.9  &  30.2 \\
 3516.944 & 46.0 & 0.9615 &  -0.21  &    2.8      &  16.2  &  14.1 \\
 3280.68  & 47.0 & 0.0000 &  -0.022 &    2.8      &  40.6  &  38.5 \\
 3382.89  & 47.0 & 0.0000 &  -0.334 &    2.8      &  29.3  &  26.9 \\
 5853.67  & 56.1 & 0.604  &  -0.91  &  0.53E-31   &  67.7  &  63.4 \\
 6141.71  & 56.1 & 0.704  &  -0.08  &  0.53E-31   & 121.2  & 114.9 \\
 6496.90  & 56.1 & 0.604  &  -0.38  &  0.53E-31   & 106.8  &  99.9 \\
 4662.50  & 57.1 & 0.0000 &  -1.24  &    2.8      &   7.6  &   6.1 \\
 4748.73  & 57.1 & 0.9265 &  -0.54  &    2.8      &   4.8  &   3.9 \\
 5303.53  & 57.1 & 0.3213 &  -1.35  &    2.8      &   4.1  &   3.2 \\
 3942.151 & 58.1 & 0.000  &  -0.22  &    2.8      &  13.4  &  11.9 \\
 3999.237 & 58.1 & 0.295  &   0.06  &    2.8      &  20.4  &  17.2 \\
 4042.581 & 58.1 & 0.495  &   0.00  &    2.8      &  12.1  &  10.0 \\
 4073.474 & 58.1 & 0.477  &   0.21  &    2.8      &  16.4  &  14.4 \\
 4364.653 & 58.1 & 0.495  &  -0.17  &    2.8      &  13.6  &  11.4 \\
 4523.075 & 58.1 & 0.516  &  -0.08  &    2.8      &  15.0  &  12.9 \\
 4562.359 & 58.1 & 0.477  &   0.21  &    2.8      &  27.4  &  24.7 \\
 5274.229 & 58.1 & 1.044  &   0.13  &    2.8      &   9.8  &   8.5 \\
 5259.73  & 59.1 & 0.633  &   0.114 &    2.8      &   2.9  &   2.3 \\
 4021.33  & 60.1 & 0.320  &  -0.10  &    2.8      &  14.8  &  12.4 \\
 4059.95  & 60.1 & 0.204  &  -0.52  &    2.8      &   7.5  &   5.4 \\
 4446.38  & 60.1 & 0.204  &  -0.35  &    2.8      &  11.9  &   9.7 \\
 5234.19  & 60.1 & 0.550  &  -0.51  &    2.8      &   6.7  &   5.2 \\
 5293.16  & 60.1 & 0.822  &   0.10  &    2.8      &  12.5  &  10.0 \\
 5319.81  & 60.1 & 0.550  &  -0.14  &    2.8      &  13.3  &  10.8 \\
 3760.70  & 62.1 & 0.18   &  -0.42  &    2.8      &   9.9  &   7.1 \\
 4318.936 & 62.1 & 0.277  &  -0.25  &    2.8      &  14.1  &  10.4 \\
 4467.341 & 62.1 & 0.659  &   0.15  &    2.8      &  14.5  &  12.6 \\
 4519.630 & 62.1 & 0.543  &  -0.35  &    2.8      &   6.7  &   5.3 \\
 4676.902 & 62.1 & 0.040  &  -0.87  &    2.8      &   6.6  &   5.0 \\
 3819.67  & 63.1 & 0.000  &   0.51  &    2.8      &  47.4  &  33.2 \\
 3907.11  & 63.1 & 0.207  &   0.17  &    2.8      &  28.8  &  19.0 \\
 4129.72  & 63.1 & 0.000  &   0.22  &    2.8      &  70.1  &  56.3 \\
 6645.10  & 63.1 & 1.379  &   0.12  &    2.8      &   8.0  &   6.6 \\
 3331.387 & 64.1 & 0.000  &  -0.28  &    2.8      &  10.2  &   7.8 \\
 3697.733 & 64.1 & 0.032  &  -0.34  &    2.8      &   6.0  &   4.5 \\
 3712.704 & 64.1 & 0.382  &   0.04  &    2.8      &  15.5  &  12.3 \\
 3768.396 & 64.1 & 0.078  &   0.21  &    2.8      &  17.7  &  14.0 \\
 4251.731 & 64.1 & 0.382  &  -0.22  &    2.8      &  13.3  &  10.4 \\
 3531.71  & 66.1 & 0.000  &   0.77  &    2.8      &  40.3  &  36.1 \\
 3536.02  & 66.1 & 0.538  &   0.53  &    2.8      &  24.4  &  19.8 \\
 3694.81  & 66.1 & 0.103  &  -0.11  &    2.8      &  17.3  &  13.9 \\
 4077.97  & 66.1 & 0.103  &  -0.04  &    2.8      &  14.8  &  10.7 \\
 4103.31  & 66.1 & 0.103  &  -0.38  &    2.8      &  17.8  &  15.0 \\
 4449.70  & 66.1 & 0.000  &  -1.03  &    2.8      &   4.3  &   3.3 \\
 3694.19  & 70.1 & 0.000  &  -0.30  &    2.8      &  81.2  &  73.9 \\
\hline                                 
\end{longtable}

\begin{table}
\begin{minipage}[t]{\textwidth}
\caption{Comparison of stellar parameters of 18 Sco}
\label{parameters}
\centering
\begin{tabular}{lrlllll} 
\hline\hline                
 \tsin & error & log $g$ & error & [Fe/H] & error & source \\
\hline    
 K     & K    & dex     & dex  &  dex   &  dex & {}     \\
\hline    
5823  &  6    & 4.45 & 0.02 & 0.054 & 0.005 & This work \\
5816  &  4    & 4.45 & 0.01 & 0.053 & 0.003 & \cite{ram14b} \\
5824  &  5    & 4.45 & 0.02 & 0.055 & 0.010 & \cite{mon13} \\
5810  & 12    & 4.46 & 0.04 & 0.05  & 0.01  & \cite{tsa13} \\
5831  & 10    & 4.46 & 0.02 & 0.06  & 0.01  & \cite{mel12} \\
5817  & 30    & 4.45 & 0.13 & 0.05  & 0.05  & \cite{das12} \\
5826  &  5    & 4.45 & 0.01 & 0.06  & 0.01  & \cite{tak09} \\
5840  & 20    & 4.45 & 0.04 & 0.07  & 0.02  & \cite{mel09}\\
5848  & 46    & 4.46 & 0.06 & 0.06  & 0.02  & \cite{ram09a}\\
5818  & 13    & 4.45 & 0.02 & 0.04  & 0.01  & \cite{sou08}\\
5834  & 36    & 4.45 & 0.05 & 0.04  & 0.02  & \cite{mel07}\\
5822  &  4    & 4.451 & 0.006 & 0.053 & 0.004 & Weighted mean from the literature\\
\hline                                 
\end{tabular}
\end{minipage}
\end{table}

\begin{table}
%\tiny
%\begin{minipage}[t]{\textwidth}
\caption{Differential abundances of 18 Sco relative to the Sun and their errors}
\label{abun}
\centering
\begin{tabular}{llrrrrlll} 
\hline\hline                
 {Element}& [X/H] & $\Delta \tef$ & $\Delta$log $g$ & $\Delta v_t$ & $\Delta$[M/H] & param\tablenotemark{a} & obs\tablenotemark{b} & total\tablenotemark{c} \\  
{}       &       & +6K           &  +0.02 dex      & +0.01 km s$^{-1}$  & +0.01 dex   &  &  &  \\
\hline
{}       & (dex) & (dex)         & (dex)           & (dex)       & (dex)        & (dex) & (dex) & (dex) \\
\hline    
C     &  -0.009 & -0.004 &  0.004 &  0.000 &  0.000 & 0.006 & 0.007 & 0.009 \\
N     &   0.003 &  0.002 &  0.004 &  0.006 &  0.007 & 0.010 & 0.013 & 0.017 \\
O     &  -0.003 & -0.005 &  0.000 & -0.001 &  0.002 & 0.005 & 0.006 & 0.008 \\
Na    &   0.025 &  0.003 &  0.000 & -0.001 &  0.000 & 0.003 & 0.003 & 0.004 \\
Mg    &   0.039 &  0.004 & -0.001 & -0.002 &  0.000 & 0.005 & 0.008 & 0.009 \\
Al    &   0.034 &  0.002 & -0.002 & -0.001 &  0.000 & 0.003 & 0.006 & 0.007 \\
Si    &   0.047 &  0.001 &  0.001 & -0.001 &  0.001 & 0.002 & 0.002 & 0.003 \\
S     &   0.016 & -0.003 &  0.003 &  0.000 &  0.001 & 0.004 & 0.007 & 0.008 \\
K     &   0.041 &  0.005 & -0.007 & -0.002 &  0.001 & 0.009 & 0.005 & 0.010 \\
Ca    &   0.057 &  0.004 & -0.003 & -0.002 &  0.000 & 0.005 & 0.003 & 0.006 \\
Sc\tablenotemark{d}    &   0.047 &  0.000 &  0.007 & -0.002 &  0.003 & 0.008 & 0.004 & 0.009 \\
Ti\tablenotemark{d}    &   0.051 &  0.006 &  0.001 & -0.001 & -0.001 & 0.006 & 0.002 & 0.007 \\
V     &   0.041 &  0.006 &  0.002 & -0.001 & -0.001 & 0.006 & 0.003 & 0.007 \\
Cr\tablenotemark{d}    &   0.056\tablenotemark{e} &  0.005 & -0.001 & -0.002 &  0.000 & 0.005 & 0.001 & 0.006 \\
Mn    &   0.041\tablenotemark{e} &  0.005 & -0.002 & -0.002 &  0.001 & 0.006 & 0.001 & 0.006 \\
Fe\tablenotemark{d}    &   0.054 &  0.004 & -0.001 & -0.002 &  0.000 & 0.005 & 0.001 & 0.005 \\
Co    &   0.027 &  0.004 &  0.002 & -0.001 &  0.000 & 0.005 & 0.003 & 0.005 \\
Ni    &   0.041 &  0.003 &  0.000 & -0.002 &  0.001 & 0.004 & 0.002 & 0.004 \\
Cu    &   0.032 &  0.003 &  0.001 & -0.002 &  0.001 & 0.004 & 0.004 & 0.006 \\
Zn    &   0.017 &  0.001 &  0.002 & -0.002 &  0.003 & 0.004 & 0.002 & 0.005 \\
Sr    &   0.097 &  0.006 &  0.000 & -0.004 & -0.001 & 0.007 & 0.011 & 0.013 \\
Y     &   0.099 &  0.001 &  0.006 & -0.005 &  0.004 & 0.009 & 0.007 & 0.011 \\
Zr    &   0.102 &  0.001 &  0.008 & -0.003 &  0.003 & 0.009 & 0.010 & 0.014 \\
Mo    &   0.115 &  0.001 &  0.002 & -0.001 &  0.000 & 0.002 & 0.015 & 0.015 \\
Ru    &   0.143 &  0.001 &  0.002 & -0.002 &  0.000 & 0.003 & 0.028 & 0.028 \\
Pd    &   0.125 &  0.001 &  0.003 & -0.002 &  0.001 & 0.004 & 0.010 & 0.011 \\
Ag    &   0.126 &  0.001 &  0.002 & -0.004 &  0.001 & 0.005 & 0.001 & 0.005 \\
Ba    &   0.118 &  0.002 &  0.002 & -0.004 &  0.005 & 0.007 & 0.006 & 0.009 \\
La    &   0.135 &  0.001 &  0.009 &  0.000 &  0.003 & 0.010 & 0.006 & 0.011 \\
Ce    &   0.117 &  0.001 &  0.009 & -0.001 &  0.003 & 0.010 & 0.006 & 0.011 \\
Pr    &   0.134 &  0.001 &  0.009 &  0.000 &  0.003 & 0.010 & 0.014 & 0.017 \\
Nd    &   0.153 &  0.002 &  0.009 & -0.001 &  0.003 & 0.010 & 0.009 & 0.013 \\
Sm    &   0.165 &  0.002 &  0.009 & -0.001 &  0.003 & 0.010 & 0.018 & 0.020 \\
Eu    &   0.187 &  0.002 &  0.007 & -0.004 &  0.004 & 0.009 & 0.030 & 0.031 \\
Gd    &   0.171 &  0.002 &  0.010 & -0.001 &  0.004 & 0.011 & 0.003 & 0.011 \\
Dy    &   0.178 &  0.002 &  0.009 & -0.002 &  0.004 & 0.010 & 0.012 & 0.016 \\
Yb    &   0.156 &  0.003 &  0.003 & -0.005 &  0.005 & 0.008 & 0.012 & 0.015 \\
\hline                                 
\end{tabular}
    \tablenotetext{a}{Errors due to stellar parameters.}
    \tablenotetext{b}{Observational errors.}
    \tablenotetext{c}{Quadric sum of errors due to observational and stellar parameter uncertainties.}
    \tablenotetext{d}{For Sc, Ti, Cr and Fe, the systematic errors due to stellar parameters
    refer to \ion{Sc}{2}, \ion{Ti}{1}, \ion{Cr}{1} and \ion{Fe}{1}, respectively.}
    \tablenotetext{e}{NLTE abundances are reported for these elements. LTE abundances are [Cr/H] = 0.058, [Mn/H] = 0.044.}
%\end{minipage}
\end{table}

\begin{table}
\begin{minipage}[t]{\textwidth}
\caption{Neutron-capture enhancement in 18 Sco}
\label{ncapture}
\centering
\begin{tabular}{lllllll} 
\hline\hline                
Z & element & [X/H]$_T$ & [X/H]$_{\rm AGB}$ & s$_{\rm Sim}$\tablenotemark{a} & s$_{\rm Bis}$\tablenotemark{b} \\
\hline    
    &     &  dex  & dex  &       &      \\
\hline    
38  & Sr  & 0.051 & 0.043 & 0.890 & 0.67  \\  
39  & Y   & 0.046 & 0.046 & 0.719 & 0.70  \\  
40  & Zr  & 0.046 & 0.056 & 0.809 & 0.64  \\  
42  & Mo  & 0.064 & 0.030 & 0.677 & 0.577 \\  
44  & Ru  & 0.094 & 0.023 & 0.39  & 0.373 \\  
46  & Pd  & 0.084 & 0.032 & 0.445 & 0.531 \\  
47  & Ag  & 0.096 & 0.013 & 0.212 & 0.221 \\  
56  & Ba  & 0.072 & 0.064 & 0.853 & 0.83  \\  
57  & La  & 0.085 & 0.055 & 0.754 & 0.711 \\  
58  & Ce  & 0.070 & 0.054 & 0.814 & 0.81  \\  
59  & Pr  & 0.084 & 0.032 & 0.492 & 0.49  \\  
60  & Nd  & 0.102 & 0.037 & 0.579 & 0.56  \\  
62  & Sm  & 0.115 & 0.018 & 0.331 & 0.31  \\  
63  & Eu  & 0.145 & 0.003 & 0.027 & 0.06  \\  
64  & Gd  & 0.118 & 0.008 & 0.181 & 0.135 \\  
66  & Dy  & 0.125 & 0.010 & 0.121 & 0.148 \\ 
70  & Yb  & 0.109 & 0.024 & 0.318 & 0.399 \\ 
\hline                                 
\end{tabular}
    \tablenotetext{a}{s-process solar system fractions by \cite{sim04}.}
    \tablenotetext{b}{s-process solar system fractions by \cite{bis11,bis13}.}
\end{minipage}
\end{table}

\end{document}